\documentclass[a4paper,fleqn]{cas-sc}
\usepackage[numbers]{natbib}
\usepackage{amssymb}
\usepackage{amsmath}
\usepackage{hyperref}
\usepackage{enumitem}
\usepackage{graphicx}%
\usepackage{multirow}%
\usepackage{amsmath,amssymb,amsfonts}%
\usepackage{amsthm}%
\usepackage{mathrsfs}%
\usepackage[title]{appendix}%
\usepackage{xcolor}%
\usepackage{textcomp}%
\usepackage{manyfoot}%
\usepackage{booktabs}%
\usepackage{listings}%
\usepackage{subfig}
\usepackage{tikz}
\usepackage{pdflscape}
\usepackage{algorithm}
\usepackage{algpseudocode}
\usepackage{float} % enables [H] placement

\def\tsc#1{\csdef{#1}{\textsc{\lowercase{#1}}\xspace}}
\tsc{WGM}
\tsc{QE}
\begin{document}
\let\WriteBookmarks\relax
\def\floatpagepagefraction{1}
\def\textpagefraction{.001}
\shorttitle{SDDPM-Polycube}
\shortauthors{}
\title [mode = title]{Scalable DDPM-Polycube: An Extended Diffusion-Based Method
  for Hexahedral Mesh and Volumetric Spline Construction}
% \author*[1]{Yuxuan Yu}\email{yuyuxuan@dhu.edu.cn}
% \author[1]{Yuzhuo Fang}\email{fangyuzhuo@mail.dhu.edu.cn}
% \author[2]{Hua Tong}\email{huat2@andrew.cmu.edu}
% \author[1]{Jiashuo Liu}\email{liujiashuo@mail.dhu.edu.cn}
% \author*[2]{Yongjie Jessica Zhang}\email{jessicaz@andrew.cmu.edu}
% \affil[1]{\orgdiv{Institute of Artificial Intelligence}, \orgname{Donghua University}, \orgaddress{\street{2999 North Renmin Road}, \city{Shanghai}, \postcode{201620}, \country{China}}}
% \affil[2]{\orgdiv{Department of Mechanical Engineering}, \orgname{Carnegie Mellon University}, \orgaddress{\street{5000 Forbes Ave}, \city{Pittsburgh}, \postcode{15213}, \state{PA}, \country{USA}}}

\author[1]{Yuxuan Yu}[orcid=0000-0001-9146-2275]
\cormark[1]
\ead{yuyuxuan@dhu.edu.cn}
\affiliation[1]{organization={Institute of Artificial Intelligence, School of
    Information and Intelligent Science, Donghua University},
  addressline={2999 North Renmin Road},
  city={Shanghai},
%          citysep={}, % Uncomment if no comma needed between city and postcode
  postcode={201620},
            country={China}}
\author[1]{Jiashuo Liu}%[]
% Footnote of the second author
% Email id of the second author
%%%%%\ead{liujiashuo@mail.dhu.edu.cn}
\author[2]{Hua Tong}%[]
% Footnote of the second author
%\fnmark[3]
% Email id of the second author
%%%%%\ead{huat2@andrew.cmu.edu}

\author[1]{Honghua Lou}%[]
% Footnote of the second author
%\fnmark[4]
% Email id of the second author
%%%%%\ead{2242795@mail.dhu.edu.cn}

\author[2]{Yongjie Jessica Zhang}[orcid=0000-0001-7436-9757]
\cormark[1]
%\fnmark[5]
\ead{jessicaz@andrew.cmu.edu}
\affiliation[2]{organization={Department of Mechanical Engineering, Carnegie Mellon University},
  addressline={5000 Forbes Ave},
  city={Pittsburgh, PA},
%          citysep={}, % Uncomment if no comma needed between city and postcode
  postcode={15213},
            country={USA}}
\cortext[1]{Corresponding author}

\begin{abstract}
  Polycube structures provide parametric domains for all-hexahedral (all-hex)
  mesh generation and analysis-suitable volumetric spline construction in
  isogeometric analysis (IGA). Recent learning-based polycube pipelines have
  improved automation, yet several challenges remain when handling complex CAD
  geometries. These challenges include the limited diversity of primitive
  geometries, restricted grid configurations, and the increasing cost of
  genus-guided context search during inference as both the primitive set and the
  grid size grow. In this paper, we present {Scalable DDPM-Polycube}, an
  extended diffusion-based polycube construction method that addresses these
  limitations. First, we expand the primitive set from two primitive geometries
  to three by introducing a blind-hole cube primitive, thereby improving the
  representation of local hole-like features that do not change the global
  genus. Second, we extend the grid configuration from the previous $2\times 1$
  setting to an enlarged three-dimensional grid configuration, which increases
  representational capacity and reduces mapping distortion for complex
  geometries. Third, we develop a genus-guided context generation strategy
  together with a hierarchical verification procedure, enabling robust context
  generation in both user-guided and automated modes. Once a valid polycube
  structure is generated, it is used for parametric mapping, all-hex control
  mesh generation, and volumetric spline construction. Experimental results
  demonstrate that scalable DDPM-Polycube improves the generality, scalability,
  and automation of diffusion-based polycube generation, and supports hex mesh
  generation and volumetric spline construction for IGA applications on complex
  geometries.
\end{abstract}
\begin{highlights}
\item Scalable DDPM-Polycube improves the scalability of diffusion-based
  polycube generation.
\item A blind-hole cube primitive and a three-dimensional grid configuration
  extend the polycube representation.
\item A genus-guided context generation strategy is developed for effective
  automated inference.
\end{highlights}
\maketitle
\section{Introduction}
Isogeometric analysis (IGA)~\cite{Hughes05a} aims to unify computer-aided design
(CAD) and finite element analysis (FEA) by using the same spline basis functions
for both geometry representation and numerical analysis. During the past two
decades, IGA has undergone substantial development. However, constructing
analysis-suitable volumetric splines from surface models remains a significant
challenge. Most industrial CAD models are represented as boundary representation
(B-Rep) models, and it is still difficult to convert B-Rep surfaces into
high-quality all-hexahedral (all-hex) control meshes and volumetric splines for
complex geometries.

In our previous CAD-to-IGA efforts with industry partners, we developed
algorithms and software for ANSYS-DYNA~\cite{yu2020hexgen,wei17a,hexdomyu}. That
pipeline semi-automatically constructs polycube structures using centroidal
Voronoi tessellation (CVT), generates all-hex control meshes through parametric
mapping and octree subdivision, converts the meshes into volumetric splines
using truncated hierarchical splines, and extracts B\'{e}zier information for
IGA. Although effective in many cases, a key difficulty is that, for B-Rep
models with complex surface features, CVT-based surface segmentation does not
always correspond to a valid polycube structure. Correction procedures are
therefore often required to satisfy geometric and topological constraints. This
issue was also emphasized in the 2023 survey by Pietroni \textit{et
  al}.~\cite{pietroni_hex-mesh_2023}. These limitations motivate a more
automated approach that can generate topology-consistent polycube structures
without case-specific heuristic operations.

Our first learning-based effort, {DL-Polycube}~\cite{yu_dl-polycube_2025},
adopts a template-mapping strategy. It predicts a polycube structure class from
an input triangular mesh and then performs surface segmentation guided by the
predicted polycube structure. This design improves automation and enables
efficient inference. Moreover, because the segmentation is driven by a valid
predicted polycube structure, it avoids the heuristic adjustments commonly
required in traditional labeling-based pipelines. However, this template-based
approach depends heavily on training coverage and on the availability of a
sufficiently rich template library. When the topology or the underlying polycube
structure lies outside the training distribution, prediction accuracy may
decrease.

Our second effort, {DDPM-Polycube}~\cite{yu_ddpm-polycube_2026}, adopts a
deformation-based learning strategy. It models the deformation from an input
geometry to its corresponding polycube structure through a denoising diffusion
process. In this pipeline, the input geometry is interpreted as a
topology-consistent polycube structure with accumulated small-scale
deformations, and the reverse diffusion process progressively removes these
deformations to recover the target polycube structure. DDPM-Polycube was
validated using two primitive geometries, namely a cube and a through-hole cube
(THC), together with a $G_{2\times 1}$ grid configuration. Experimental results
showed that DDPM-Polycube generates polycube structures for models up to genus 2
and can also handle cases whose topology lies outside its training range. These
results support the idea that learning deformation is more scalable than
explicitly enumerating large template libraries, as in
DL-Polycube. Nevertheless, DDPM-Polycube still has three important limitations
that must be addressed before it can become a more practical and fully automated
CAD-to-IGA solution.

The first limitation is the restricted primitive set. In DDPM-Polycube, only two
primitive geometries are used: a genus-0 cube and a genus-1 THC. In real CAD
models, however, hole-like features are not always reflected by the global
genus. A typical example is a blind-hole feature, which starts from an exterior
surface and terminates inside the solid without exiting elsewhere. Such a
feature remains genus-0, even though it exhibits a local hole-like
structure. This differs from a through-hole feature, in which the hole passes
completely through the solid and changes the genus. If a blind-hole feature is
represented only by combining a genus-0 cube and a genus-1 THC, the inference
process becomes ambiguous, because the global genus alone is not sufficient to
characterize the geometric composition.

The second limitation is the restricted grid configuration. The original
DDPM-Polycube was mainly validated on the simplest configuration,
$G_{2\times 1}$, which is effectively a one-dimensional (1D) grid with at most
two active cells. Although this setting is sufficient as a proof of concept, it
is not adequate for many engineering geometries. Forcing a complex geometry into
only two cells often causes large mapping distortion and stretched hex elements,
which in turn degrades mesh quality and affects subsequent volumetric spline
construction. Therefore, a richer three-dimensional (3D) grid configuration is
needed.

The third limitation lies in scalable context generation during
inference. DDPM-Polycube uses a context vector to condition the reverse
diffusion process, and candidate contexts are filtered using the genus of the
input triangular mesh. In the $2\times 1$ setting, the number of candidates
remains manageable, so one can traverse the candidate set, run reverse diffusion
for each candidate, and accept a valid result after verification. However, this
strategy does not scale well. When the primitive set grows and the grid becomes
larger, the number of genus-consistent candidates increases rapidly. Inference
then becomes dominated by context traversal and candidate verification, leading
to substantially higher computational cost.

To address these limitations, we extend DDPM-Polycube and develop {Scalable
  DDPM-Polycube} (SDDPM). This paper makes the following three main
contributions:
\begin{enumerate}
\item We expand the primitive set from two geometries to three by introducing a
  blind-hole cube (BHC) primitive. This expansion addresses an important limitation of
  DDPM-Polycube under genus-guided conditioning. Some geometries contain local
  hole-like structures that are not uniquely reflected by the global genus. With
  the added primitive, such cases can be characterized more consistently during
  primitive geometry assembly.
\item We move beyond the 1D $G_{2\times 1}$ setting and extend the pipeline to a
  higher-dimensional 3D grid configuration. This enables richer polycube
  structures, improves representational capacity, and reduces mapping distortion
  in subsequent all-hex control mesh generation.
\item We introduce a genus-guided context generation algorithm for scalable
  inference. The algorithm supports both user-guided and automated modes. In the
  automated mode, it constructs global contexts from locally inferred subregions
  under genus-based constraints and applies a two-stage hierarchical
  verification module to discard invalid candidates.

\end{enumerate}

The remainder of this paper is organized as follows. Section 2 reviews related
work. Section 3 provides an overview of the SDDPM
pipeline. Section 4 describes dataset generation, context encoding, and feature
extraction. Section 5 presents the diffusion model architecture. Section 6
introduces the genus-guided context generation algorithm and the hierarchical
verification module. Section 7 presents experimental results and
evaluations. Finally, Section 8 concludes the paper and discusses future work.

\section{Related work}
\label{sec:related-works}

A variety of approaches to all-hex mesh generation, learning-based mesh generation,
and volumetric spline construction have been developed in recent years. Since
SDDPM builds upon these research directions, we briefly review prior work from
the following three aspects. Together, they provide the foundation for the
proposed pipeline.

\subsection{All-hex control meshes for IGA}

IGA requires volumetric spline construction whose control meshes are typically
discretized using tetrahedral or hex elements. Compared with tetrahedral meshes,
hex meshes are often preferred because they can achieve comparable accuracy with
fewer elements~\cite{benzley1995comparison}, reduce locking
effects~\cite{francu2021locking}, and naturally support tensor-product spline
construction. Despite extensive progress in all-hex mesh
generation~\cite{pietroni_hex-mesh_2023,ref:zhangbook,zhang2013challenges},
generating high-quality all-hex meshes for complex B-Rep models remains
challenging.

A variety of strategies have been explored for all-hex mesh generation, including
indirect methods~\cite{E96}, sweeping
methods~\cite{Zhang20072943,yu2019anatomically}, grid-based
methods~\cite{qian2012automatic,schneiders1996grid,qian2010quality,ZhangJ2006},
polycube
methods~\cite{Tarini2004,Wang07polycubesplines,wang2013trivariate,HZ2015CMAME,yu2020hexgen,yu_dl-polycube_2025,yu_ddpm-polycube_2026},
and vector field-based
methods~\cite{nieser2011cubecover,Li2012AMU,bruckler2024integer,bruckler2022volume}. A
major issue is that all-hex meshes may contain extraordinary edges and extraordinary
points. In IGA, a high density of such extraordinary features may make it
difficult to obtain optimal convergence rates. Therefore, methods that can
generate meshes with as few extraordinary edges and extraordinary points as
possible are generally preferred.

In this context, sweeping methods and polycube methods are particularly
relevant. Sweeping methods generate all-hex meshes by propagating a mesh from source
to target surfaces, but they usually require compatible topologies between the
swept boundaries, which limits their applicability. Polycube methods introduce a
structured parameter space in each individual cube and generate all-hex meshes
through mapping. The polycube concept was initially proposed for texture
mapping~\cite{Tarini2004} and was later adopted in all-hex mesh generation
pipelines. Lin \textit{et al}.~\cite{LinJ2008} presented an automated method for
constructing polycubes, but it is less robust for models with complex
topology. Huang \textit{et al}.~\cite{huang2014} introduced an $\ell_1$-based
deformation energy defined on a tetrahedralization to construct polycubes. Fu
\textit{et al}.~\cite{fu2016efficient} further introduced an AMIPS term to
improve distortion control and avoid degenerate or inverted elements. Li
\textit{et al}.~\cite{li2021interactive} proposed an interactive pipeline that
incorporates user input to improve the quality of the generated polycube
structures.

An important advantage of polycube structures is that they provide direct
control over the number and placement of extraordinary edges and extraordinary
points, which supports IGA-suitable hex control meshes and subsequent spline
construction. For complex B-Rep models, generalized polycube methods further
expand the adaptability of polycube pipelines to high-genus and complex
geometries~\cite{LiBo2012}. Related efforts include CVT-driven surface
segmentation for defining polycube
structures~\cite{HZ2015CMAME,HZL2016,yu2020hexgen}, modifying polycube
structures by introducing extraordinary edges to improve mesh
quality~\cite{guo_cut-enhanced_2020}, and extensions that generate all-hex meshes
with structures free of interior extraordinary
points~\cite{fang2016all,mandad2022intrinsic}. However, for B-Rep models with
complex surface features, surface labeling does not always lead to a valid
polycube structure~\cite{pietroni_hex-mesh_2023}. Practical pipelines therefore
often require correction procedures to satisfy geometric and topological
constraints. These limitations motivate more automated methods that can produce
topology-consistent polycube structures without relying on case-by-case
heuristic operations.
\subsection{Learning-based mesh generation}
Machine learning has recently become a powerful tool for automating and
improving mesh generation pipelines. Transformer-based models have been applied
to surface mesh generation~\cite{Siddiqui2024}, and learning-based frameworks
have been developed for tetrahedral reconstruction~\cite{Gao2020} as well as
triangular mesh generation from point clouds~\cite{Chen2024}. Beyond surface
meshes, neural methods have also been integrated with classical meshing
strategies. Tong \textit{et al}.~\cite{TONG2023102109} combined neural networks with
advancing-front techniques to generate planar unstructured quadrilateral meshes.

For IGA-oriented all-hex control mesh generation, Yu \textit{et
  al}.~\cite{yu_dl-polycube_2025} integrated deep learning with the polycube
method by directly predicting polycube structures and then generating hex
control meshes through parametric mapping. This data-driven pipeline reduces
manual heuristic adjustments in labeling by learning a template-driven mapping
from geometry to polycube structures. However, its performance is constrained by
template coverage. Covering diverse design parameters and topologies requires a
large collection of representative polycube structures, whose enumeration is
costly and does not scale well.

Generative modeling offers an alternative to explicit template enumeration by
learning probability distributions over shapes. Common families of generative
models include autoregressive sequence models~\cite{sutskever2014sequence},
generative adversarial networks~\cite{goodfellow2014generative}, flow-based
models~\cite{dinh2016density,kingma2018glow}, variational
autoencoders~\cite{kingma2013auto}, and diffusion
models~\cite{sohl2015deep,ho2020denoising}. Among them, diffusion models have
demonstrated strong sample quality and stable training in high-dimensional
generation tasks. Denoising diffusion probabilistic models (DDPM) learn a
reverse-time process that progressively removes noise~\cite{ho2020denoising},
while score-based formulations interpret sampling as solving stochastic
differential equations~\cite{song2020score}. Advances in sampling and denoising
strategies further improve robustness to noise and
outliers~\cite{nichol2021improved}. Diffusion models have also been applied in
computational mechanics~\cite{KARTASHOV2025117742}, suggesting their potential
beyond image synthesis.

Yu \textit{et al}.~\cite{yu_ddpm-polycube_2026} proposed DDPM-Polycube, which
learns the deformation from an input geometry to a topology-consistent polycube
structure using a modified DDPM formulation. In this formulation, the input
geometry is interpreted as a polycube structure with accumulated small-scale
stochastic deformations, and reverse diffusion progressively removes these
deformations to recover a parameter space suitable for polycube-based hex
meshing. By learning deformation rather than relying on large libraries of
predefined templates and explicit template mappings, DDPM-Polycube improves
generalization under limited training data and demonstrates effectiveness on
geometries up to genus 2. However, DDPM-Polycube is still limited by the scope
of its primitive set and grid configuration, and its inference stage uses a
genus-guided traversal of candidate context vectors. When the primitive set
expands and the grid becomes larger, the number of candidate contexts can grow
rapidly, leading to higher inference cost.

\subsection{Volumetric spline construction}
Once a hex control mesh is available, constructing an analysis-suitable
volumetric spline remains a key challenge. Volumetric parameterization
techniques include NURBS~\cite{Zhang20072943},
T-splines~\cite{Wang07polycubesplines,zhang_solid_2012,wang2013trivariate}, and
TH-splines~\cite{wei17a}. T-splines enable local refinement, while TH-splines
use truncation to reduce basis overlap and improve numerical conditioning. Local
refinement and fitting strategies also play an important role in improving
geometric accuracy~\cite{LiBo2012}. These developments provide a suitable
foundation for combining polycube-based all-hex control meshing with volumetric spline
construction in IGA.

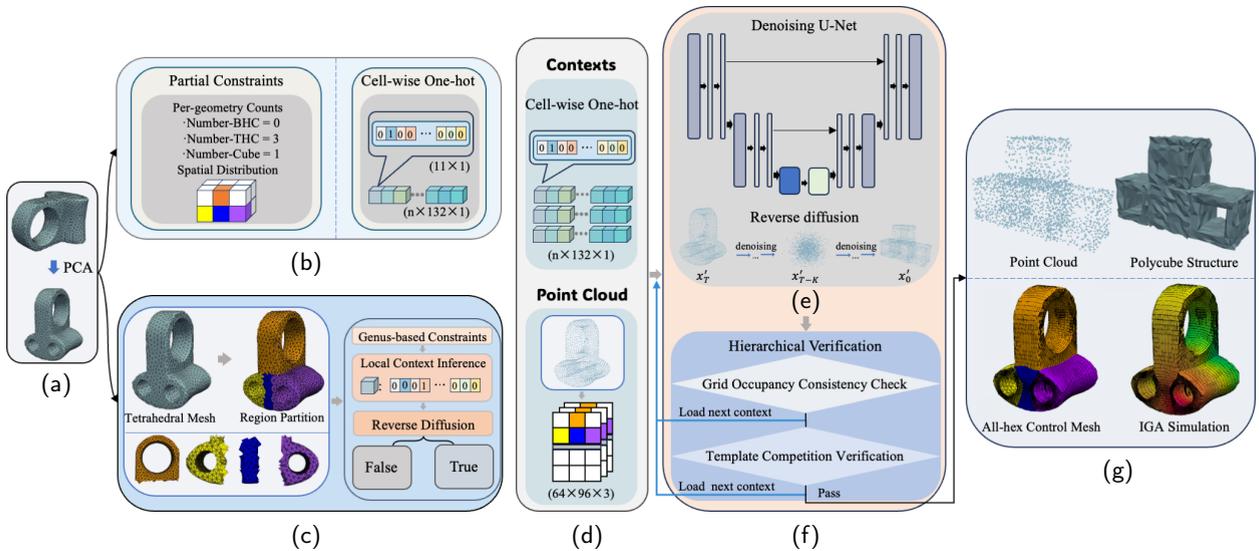
\begin{figure}[pos=htbp]
      \centering
  \begin{tikzpicture}
    \node[anchor=south west,inner sep=0] (image) at (0,0) {\includegraphics[width=\linewidth]{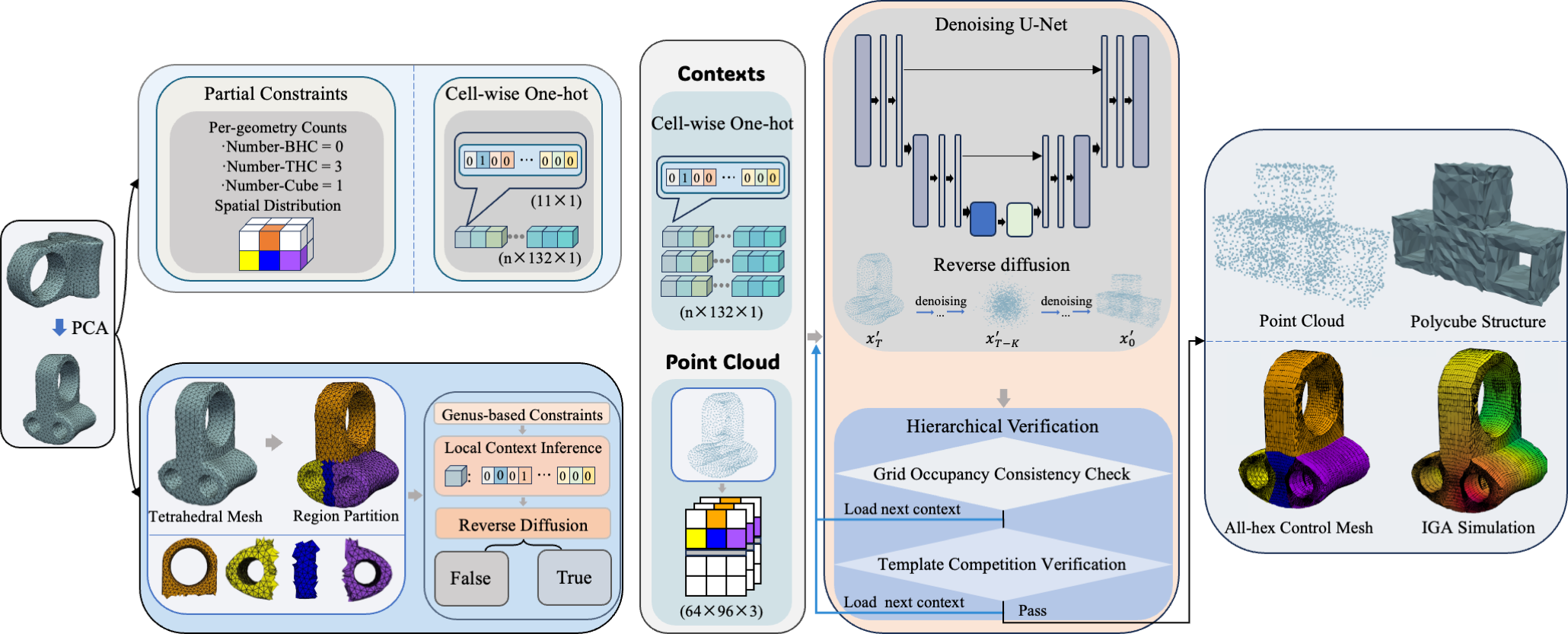}};
    \begin{scope}[x={(image.south east)},y={(image.north west)}]
      \node at (0.04,0.25) {\small (a)};
      \node at (0.24,0.49) {\small (b)};
      \node at (0.24,-0.05) {\small (c)};
      \node at (0.465,-0.05) {\small (d)};
      \node at (0.64,0.415) {\small (e)};
      \node at (0.64,-0.05) {\small (f)};
      \node at (0.89,0.08) {\small (g)};
    \end{scope}
  \end{tikzpicture}
  \caption{\label{fig:Pipeline_Structure_V2} Overview of the SDDPM pipeline.
    (a) The input triangular mesh is preprocessed by PCA alignment, centering,
    and normalization.  (b) In the user-guided mode, partial constraints or one
    or more complete cell-wise one-hot encoded context vectors are provided and
    converted into candidate global contexts when needed.  (c) In the automated
    mode, the input geometry is temporarily partitioned into subregions, local
    contexts are inferred under genus-based constraints, and verified local
    results are assembled into a global context.  (d) The candidate contexts are
    paired with the preprocessed geometry, which is converted into a point cloud
    and reshaped into a \(64\times 96\times 3\) tensor.  (e) The denoising U-Net
    performs context-conditioned reverse diffusion to generate a candidate
    polycube structure.  (f) The candidate is validated by hierarchical
    verification through the grid occupancy consistency check and the template
    competition verification, with failed candidates routed to the next context.
    (g) The accepted polycube structure is used for parametric mapping, all-hex
    mesh generation, mesh quality improvement, volumetric spline construction,
    and IGA.}
\end{figure}

\section{Overview of the SDDPM pipeline}
\label{sec:pipeline_overview_pp}

Fig.~\ref{fig:Pipeline_Structure_V2} shows the overall pipeline of SDDPM, which
converts an input triangular mesh into a topology-consistent polycube structure,
and then further generates an all-hex control mesh and volumetric splines for
IGA. The pipeline contains seven components, shown in
Fig.~\ref{fig:Pipeline_Structure_V2}(a)--(g), and supports two inference modes,
a user-guided mode and an automated mode. The pipeline starts from the input
triangular mesh and a preprocessing step, as shown in
Fig.~\ref{fig:Pipeline_Structure_V2}(a). We first apply principal component
analysis (PCA) to align the geometry into a consistent reference
orientation. After PCA alignment, the model is centered and normalized so that
its scale is compatible with the training data. This preprocessing step is
important because the subsequent context generation and reverse diffusion are
performed under a fixed grid-based representation.

After preprocessing, the pipeline proceeds in one of two ways, depending on
whether user guidance is available. In the user-guided mode
(Fig.~\ref{fig:Pipeline_Structure_V2}(b)), the user may provide either partial
constraints or one or more complete cell-wise one-hot encoded context
vectors. The partial constraints are specified in terms of per-geometry counts
and spatial distribution. As shown in the figure, the user can prescribe the
numbers of different primitive geometries, such as \texttt{Number-BHC}
(blind-hole cube), \texttt{Number-THC} (through-hole cube), and
\texttt{Number-Cube}, together with a coarse spatial arrangement of these
primitives on the grid.  These partial constraints are then converted into
cell-wise one-hot encoded context vectors. For each cell, the local context is
an 11-dimensional one-hot vector, shown as \((11\times 1)\) in
Fig.~\ref{fig:Pipeline_Structure_V2}(b), corresponding to 10 primitive-category
states plus one null state. By concatenating the local vectors over all 12
cells, we obtain a global context vector of size \((132\times 1)\), also shown
in Fig.~\ref{fig:Pipeline_Structure_V2}(b). If the user directly provides one or
more complete cell-wise one-hot encoded context vectors, this conversion step is
not needed. When multiple such context vectors are available, they are stacked
as \((n\times 132\times 1)\), where \(n\) denotes the number of context vectors
considered during inference so that different candidate results can be examined.

If no user guidance is given, the pipeline enters the automated mode, shown in
Fig.~\ref{fig:Pipeline_Structure_V2}(c). In this mode, we first construct a
temporary volumetric representation of the geometry and partition it into
subregions. The figure shows both the volumetric partition of the input model
and the corresponding extracted boundary surfaces of these subregions. For each
subregion, we compute its genus and use genus-based constraints to reduce the
locally feasible primitive categories. Because genus alone is not sufficient to
determine a unique primitive label, we then perform local context inference for
each subregion. The inferred local candidate is evaluated through reverse
diffusion and hierarchical verification. If the local candidate fails, it is
rejected. If it succeeds, it is accepted and used to assemble the final global
context. In this way, the automated mode avoids direct traversal of the full
global context space and instead constructs the global context from verified
local inferences.

The resulting context information, regardless of whether it comes from the
user-guided mode or the automated mode, is organized as the context module
shown in Fig.~\ref{fig:Pipeline_Structure_V2}(d).  In this module, the candidate
contexts are paired with the preprocessed input geometry for context-conditioned
reverse diffusion. The preprocessed input geometry is converted into a point
cloud representation and then reshaped into a three-channel tensor of size
\((64\times 96\times 3)\), also shown in
Fig.~\ref{fig:Pipeline_Structure_V2}(d). This tensor is the image-like input
used by the diffusion model, where the three channels store the spatial
coordinates of the point cloud.

Given the context vector and the point-cloud tensor, the pipeline performs
context-conditioned reverse diffusion using the denoising U-Net, as shown in
Fig.~\ref{fig:Pipeline_Structure_V2}(e). Starting from the input geometry
\(\mathbf{x}'_T\), the model progressively removes small-scale deformations
through the reverse diffusion process and generates a polycube structure
candidate \(\mathbf{x}'_0\). The arrows in
Fig.~\ref{fig:Pipeline_Structure_V2}(e) indicate the iterative denoising steps,
and the U-Net architecture provides the denoising prediction at each timestep.
This stage is the core generative component of SDDPM. The generated polycube
candidate is then passed to the hierarchical verification module in
Fig.~\ref{fig:Pipeline_Structure_V2}(f). This module contains two stages. The
first stage is the grid occupancy consistency check (GOCC), which examines
whether the generated point distribution is consistent with the context in terms
of occupied cells and coarse spatial extent. The second stage is the template
competition verification (TCV), which further checks whether the generated
geometry in each occupied cell matches the intended primitive category and
orientation. If a candidate fails either stage, the pipeline loads the next
context as shown in Fig.~\ref{fig:Pipeline_Structure_V2}(f) and tests the next
candidate context. If the candidate passes both stages, the pipeline follows the
pass branch and accepts the polycube structure.

As shown in Fig.~\ref{fig:Pipeline_Structure_V2}(g), once a polycube structure
is accepted, it is used for subsequent meshing steps. The pipeline first
constructs a bijective surface mapping between the input triangular surface and
the generated polycube structure. It then uses the generated polycube structure
as the parametric domain to generate an all-hex control mesh by parametric
mapping~\cite{zhang_solid_2012} combined with octree subdivision. The initial mesh
is then improved by mesh quality operations, including
pillowing~\cite{YZhang2009c}, smoothing, and Jacobian-based
optimization~\cite{tong2026hexopt,qian2012automatic,tong_hybridoctree_hex_2024,tong2024fast}.
Mesh quality is quantified using the scaled Jacobian. We use the minimum scaled
Jacobian after optimization in the physical domain as the primary quality
indicator. If the resulting mesh quality is not satisfactory, the pipeline
continues to test additional candidate contexts. It generates the corresponding
polycube structures and repeats parametric mapping, all-hex control mesh generation, and
mesh quality improvement. This iterative process continues until a satisfactory
mesh quality is obtained. Finally, based on the optimized all-hex control mesh, we
construct analysis-suitable volumetric splines using TH-spline3D with local
refinement~\cite{wei17a,yu2020hexgen}. We then extract trivariate B\'{e}zier
information and transfer it to ANSYS-DYNA for IGA.

\section{Dataset generation, context encoding, and feature extraction}
\label{sec:data_context_feature}

This section presents the data generation and context encoding used to train
SDDPM with an expanded primitive set and a higher-dimensional
grid configuration. The objective is to construct a structured dataset so that
the model learns deformation priors from a small set of primitive geometries and
their assemblies, while the inference process can be conditioned by an explicit
genus-guided context.

\subsection{Expanded primitive set and orientation variants}
\label{sec:expand-prim-set}
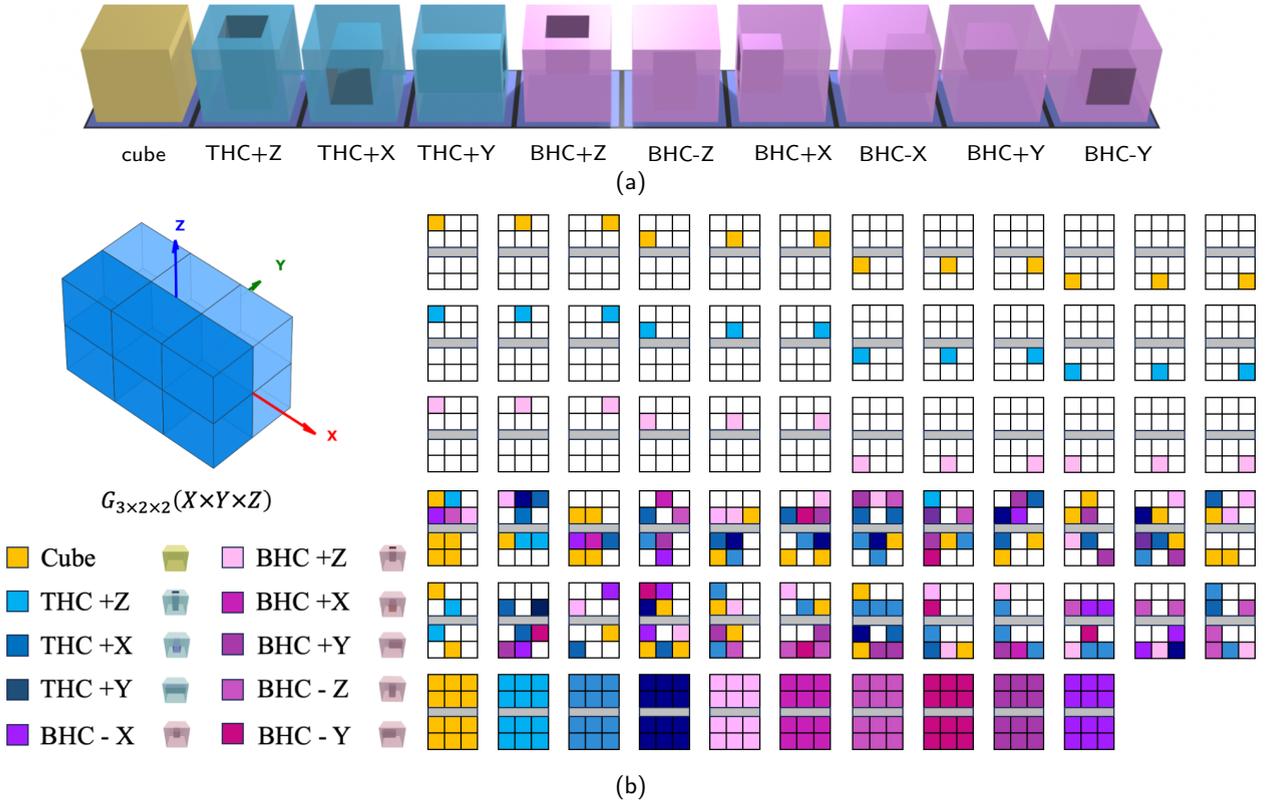
\begin{figure}[pos=htbp]
  \centering
  \begin{tikzpicture}
    \node[anchor=south west,inner sep=0] (image) at (0,0) {\includegraphics[width=\linewidth]{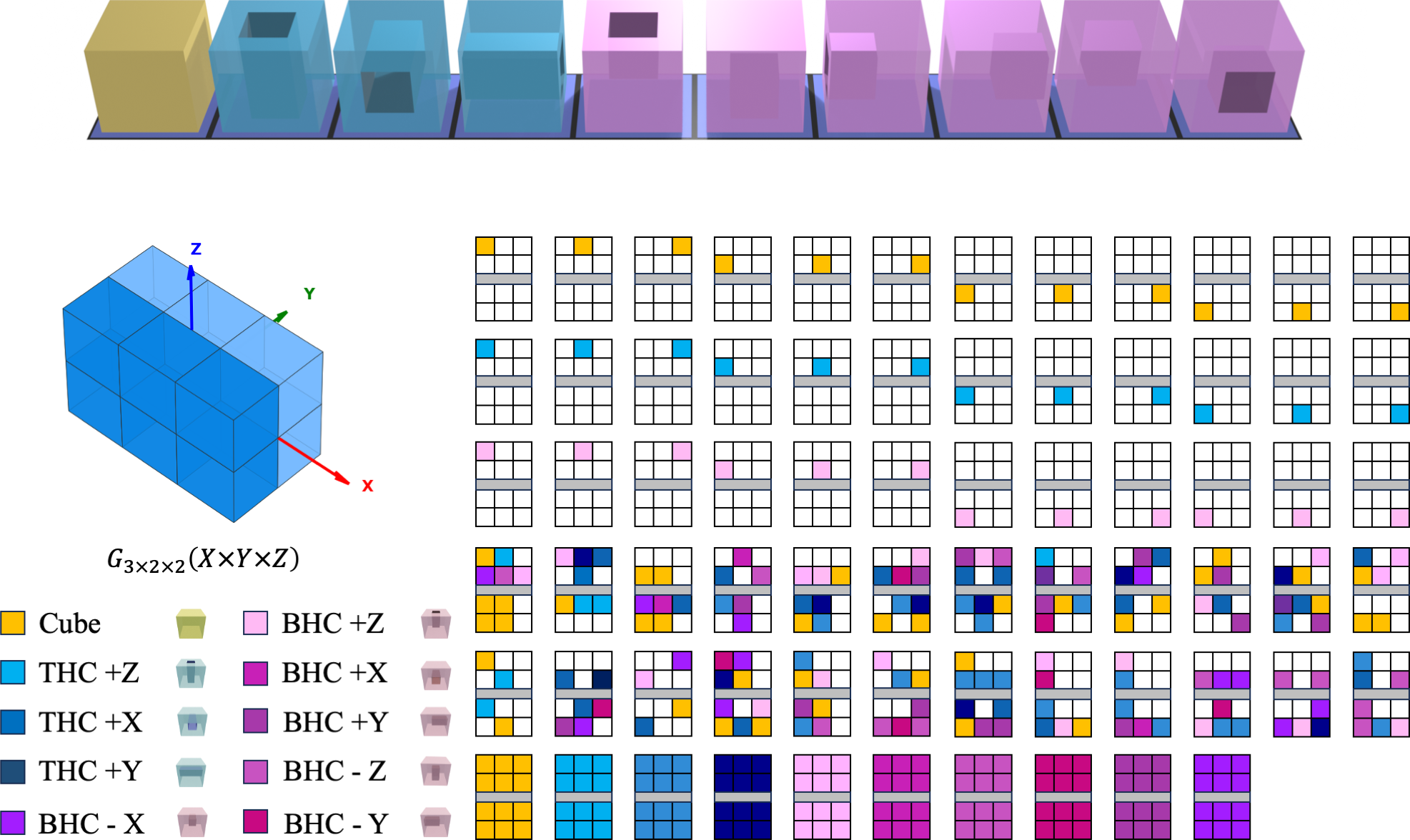}};
    \begin{scope}[x={(image.south east)},y={(image.north west)}]
      \node at (0.11,0.8) {\footnotesize cube};
      \node at (0.19,0.8) {\footnotesize THC+Z};
      \node at (0.28,0.8) {\footnotesize THC+X};
      \node at (0.36,0.8) {\footnotesize THC+Y};
      \node at (0.45,0.8) {\footnotesize BHC+Z};
      \node at (0.54,0.8) {\footnotesize BHC-Z};
      \node at (0.63,0.8) {\footnotesize BHC+X};
      \node at (0.71,0.8) {\footnotesize BHC-X};
      \node at (0.8,0.8) {\footnotesize BHC+Y};
      \node at (0.89,0.8) {\footnotesize BHC-Y};
      \node at (0.5,0.76) {\small (a)};
      \node at (0.5,-0.05) {\small (b)};
    \end{scope}
  \end{tikzpicture}
  \caption{\label{fig:grid_unfold}Expanded primitive set and unfolded grid
    layout used for dataset generation and context encoding. (a) The primitive
    library contains ten axis-aligned categories derived from three base
    primitive geometries: a cube, a THC with three axial orientations, and a BHC
    with six directional variants. (b) These primitives are assembled on the
    $G_{3\times 2\times 2}$ grid, which is unfolded into a fixed $4\times 3$
    layout for data generation. Representative examples of single-cell, randomly
    sampled multi-cell, and full-grid assemblies are also shown.}
\end{figure}

We define a primitive set for polycube construction that contains three
primitive geometries, as shown in Fig.~\ref{fig:grid_unfold}(a): a cube
(genus-0), a THC (genus-1), and a BHC (genus-0). The THC models a through
feature that changes the global genus, whereas the BHC models a local hole-like
feature that does not change the global genus. Each primitive is generated as a
triangular mesh and then sampled into a point cloud. To support
orientation-aware conditioning, we include axis-oriented variants for the THC
and the BHC. The THC is represented by three variants whose hole axis is aligned
with the $Z$, $X$, or $Y$ axis. The BHC is represented by six variants whose
blind-hole direction is aligned with one of the $\pm Z$, $\pm X$, or $\pm Y$
directions. The resulting primitive category set is
\begin{equation}
  \label{eq:category_set}
  \mathcal{K}=
  \{\ \texttt{cube},\ \texttt{THC+}Z,\ \texttt{THC+}X,\ \texttt{THC+}Y,\
  \texttt{BHC+}Z,\ \texttt{BHC-}Z,\ \texttt{BHC+}X,\ \texttt{BHC-}X,\ \texttt{BHC+}Y,\ \texttt{BHC-}Y\}.
\end{equation}
Here, the ``$\pm$'' sign in the BHC categories distinguishes the two directional
variants along the same axis.

\subsection{Grid configuration and representative assemblies}
\label{sec:grid_config}
We represent a polycube structure on a discrete 3D grid configuration
$G_{3\times 2\times 2}$ with 12 cells. Each cell stores either one primitive
category in $\mathcal{K}$ or a null state indicating an empty location. This
$3\times 2\times 2$ setting is used as the main configuration for both data
generation and inference. For visualization and implementation convenience, we
unfold $G_{3\times 2\times 2}$ into a $4\times 3$ 2D layout with two slices, as
shown in Fig.~\ref{fig:grid_unfold}(b). This unfolding provides a fixed layout
of 12 cell positions that is used consistently in both data generation and
inference, so that each position always corresponds to the same cell in the
original 3D grid.

On this grid, representative assemblies are constructed by placing primitives
into occupied cells. For single-cell instances, each primitive category can be
placed in any of the 12 cell locations, resulting in $12\times 10=120$
single-cell configurations. To relate our dataset construction to that in
DDPM-Polycube~\cite{yu_ddpm-polycube_2026}, we consider the special case in
which only the cube primitive is used and the occupied-cell set is allowed to
vary. In this case, the number of distinct multi-cell assemblies is determined
purely by combinatorics. With 12 cells, the number of occupancy patterns with at
least two occupied cells is $\sum_{k=2}^{12}\binom{12}{k}=4,083$. In the
DDPM-Polycube setting, including such combinations in the training set helps the
model learn occupancy and assembly. However, if multi-cell examples are
dominated by cube while other primitive categories appear mainly in single-cell
or fully occupied cases, the learned prior may become overly sensitive to cube
patterns and suppress other primitives during inference. This motivates a
dataset construction strategy that simultaneously covers the combinatorial
occupancy space and maintains primitive diversity.

In our implementation, the training set is generated using three components, as
illustrated in Fig.~\ref{fig:grid_unfold}. First, we include all single-cell
configurations (120 samples in total). This ensures that each primitive category
in $\mathcal{K}$ appears at every cell location while all other cells are set to
null. Second, we include full-grid configurations by occupying all 12 cells with
the same primitive category. This provides fully occupied reference cases for
each category. Third, we generate 4,870 additional assemblies by balanced random
sampling over occupancy counts. The number of occupied cells is chosen from 1 to
11 with an approximately uniform distribution. For each sample, we randomly
select that many active cells from the 12 cells, assign to each active cell a
primitive category uniformly at random from $\mathcal{K}$, and set all inactive
cells to null. During geometry generation, we apply a small random isotropic
scaling to the point coordinates by multiplying the point cloud by a factor
sampled from the uniform distribution $U(0.99,1.01)$. Unless otherwise stated,
the random number generator seed is fixed to 42 for reproducibility.
Fig.~\ref{fig:grid_unfold} presents representative examples from these three
components, including single-cell configurations, full-grid configurations, and
randomly generated multi-cell assemblies.

\subsection{Context encoding}
\label{sec:context_encoding}
SDDPM uses a fixed-length cell-wise one-hot context vector to represent the
cell-wise assembly on the grid $G_{3\times 2\times 2}$. The grid contains 12
cells, and each cell is either empty or occupied by one primitive category in
$\mathcal{K}$. Since $|\mathcal{K}|=10$, each cell has 11 possible states,
including the empty state null. The context is therefore a 132-dimensional
binary vector, where $132=12\times 11$. It is formed by concatenating the
one-hot encoded context vectors corresponding to the 12 cells. We denote the
context by
$\mathbf{c}=[\mathbf{c}^{(1)};\ldots;\mathbf{c}^{(12)}]\in\{0,1\}^{132}$, where
$\mathbf{c}^{(i)}\in\{0,1\}^{11}$ encodes the state of cell $i$.

We index the 11 states by $m\in\{0,\ldots,10\}$ and reserve $m=0$ for the null
state. For each cell $i$, we define a discrete label $\ell_i\in\{0,\ldots,10\}$
and encode it by $c_m^{(i)}=\delta_{m,\ell_i}$. The label $\ell_i$ is obtained
from occupancy and primitive category assignment. Let $o_i\in\{0,1\}$ denote
whether cell $i$ is occupied. If $o_i=0$, we set $\ell_i=0$. If $o_i=1$, we
assign one primitive category $k_i\in\mathcal{K}$ to cell $i$ and map it to an
index in $\{1,\ldots,10\}$ using a fixed ordering of $\mathcal{K}$ shared by all
samples. The resulting 132-dimensional vector $\mathbf{c}$ is the cell-wise
assembly encoding of the sample. It is known when we assemble the training
geometry and remains unchanged for all diffusion timesteps.

During training, we first construct the assembled geometry $\mathbf{x}_0$ using
the primitive set and the grid-based assembly procedure. The context vector
$\mathbf{c}$ is created from the same assembly information. We then apply the
forward diffusion process to obtain $\mathbf{x}_t$ while keeping the same
context $\mathbf{c}$, and train a context-conditioned denoising model with
inputs $(\mathbf{x}_t,t,\mathbf{c})$. At inference, the same context interface
is used to condition the reverse diffusion process.

\subsection{Feature extraction}
\label{sec:feature_extraction}
Once the dataset is generated, the next step is to convert each geometric
configuration into a representation suitable for the SDDPM
model. In our pipeline, each training sample is an assembled polycube-like
geometry on the grid $G_{3\times 2\times 2}$ with 12 cells. We first transform
the triangular mesh representation of the assembled geometry into a point cloud
and then map it into an image-like tensor that can be processed by the U-Net
architecture.

Let the sampled point cloud be \(P=\{p_1,p_2,\ldots,p_N\}\), where each point is
\(p_i=(x_i,y_i,z_i)\in\mathbb{R}^3\), and $N$ is the total number of sampled
points. Let $N_c$ denote the fixed number of sampled points per cell. Since our
grid contains 12 cells and we allocate $N_c=512$ points to each cell, we set
$N=12N_c=6,144$. Points are sampled on the triangular surface of the primitive
placed in each occupied cell. If a cell is null, it contributes no surface
points. To keep the input size fixed for diffusion training, we insert a
placeholder point set for each empty cell, defined as the constant point
coordinate $\mathbf{0}$ repeated $N_c$ times. This ensures that all samples have
the same dimensionality and that the cell-wise layout remains consistent.

To ensure scale consistency across the dataset, we normalize the point
coordinates to the range $[-1,1]$. In practice, we compute the axis-aligned
bounding box of the assembled geometry, apply an affine scaling so that the
longest box edge maps to length 2, and translate the box center to the origin. A
key design objective of SDDPM is to preserve grid context during
learning. Therefore, we impose a deterministic cell-wise ordering consistent
with the unfolded $4\times 3$ layout. We index the 12 cells by
$i\in\{1,\ldots,12\}$ following this fixed layout. For each cell $i$, we sample
512 points and sort them by the $X$, $Y$, and $Z$ coordinates in ascending
order. We then concatenate the cells in index order to obtain a globally ordered
list of $N$ points. This ordering is used in both training and inference so that
the model always observes the same tensor region for the same grid cell.

To adapt the point cloud for the DDPM framework, we reorganize the $N$ points
into an image-like tensor by enforcing a fixed blockwise layout consistent with
the unfolded $4\times 3$ grid in Fig.~\ref{fig:grid_unfold}. Specifically, the
ordered point set is partitioned into $4\times 3$ blocks, resulting in 12
blocks, each of size $16\times 32$. Each block corresponds to one grid cell and
stores the points sampled from that cell. For each cell, we sort its local point
set according to the $X$, $Y$, and $Z$ coordinates and then write the sorted
points into the corresponding $16\times 32$ block in row-major order. This
blockwise filling enforces a strict correspondence between matrix subregions and
geometric components in different cells, which strengthens the model’s ability
to learn multi-part spatial layouts under the grid-based context. Based on this
reshaping, we organize the geometry into a three-channel tensor
$X\in\mathbb{R}^{64\times 96\times 3}$, where the channel dimension stores the
original $X$, $Y$, and $Z$ coordinates.

During inference, the input geometry is treated as $\mathbf{x}'_T$ and is
normalized using the same rule as the training data. It is then converted into a
three-channel $64\times 96$ tensor using the same cell-wise ordering, sorting,
and reshaping procedure, and provided to the reverse diffusion process together
with the timestep embedding and the context vector $\mathbf{c}$. After reverse
diffusion, the output tensor is decoded back into 3D point coordinates. The
fixed ordering guarantees that the generated point set preserves the same point
count and cell-wise layout as the input representation.

\section{SDDPM model architecture}
\label{sec:ddpm-polycubepp-arch}
This section summarizes the SDDPM model architecture. The proposed model follows
the DDPM-Polycube formulation~\cite{yu_ddpm-polycube_2026}. For completeness, we
briefly review the essential equations and implementation details needed for
reproducibility. The main differences from the original DDPM-Polycube setting
lie in the enlarged input resolution and the higher-dimensional context encoding
associated with the $G_{3\times 2\times 2}$ grid.

Given an input geometry tensor
$\mathbf{x}'_T\in\mathbb{R}^{64\times 96\times 3}$
(Section.~\ref{sec:feature_extraction}) and a grid-aware context vector
$\mathbf{c}\in\{0,1\}^{132}$ (Section.~\ref{sec:context_encoding}), SDDPM
generates a topology-consistent polycube structure $\mathbf{x}'_0$ through a
context-conditioned diffusion model. As in
DDPM-Polycube~\cite{yu_ddpm-polycube_2026}, we interpret the input shape as a
polycube structure with accumulated small-scale deformations and recover the
polycube by iteratively removing these deformations through reverse diffusion.

The forward diffusion process progressively deforms a polycube structure
$\mathbf{x}_0$ by introducing Gaussian noise at each timestep. Following
DDPM-Polycube, we use a nonzero-mean Gaussian noise $\mathcal{N}(\mathbf{q},I)$
so that the endpoint of the forward process remains a geometry rather than
converging to a standard Gaussian distribution. Let $\alpha_t=1-\beta_t$ and
$\bar{\alpha}_t=\prod_{i=1}^t\alpha_i$, where $\{\beta_t\}_{t=1}^T$ is a
predefined schedule. Using the reparameterization trick, the forward process
admits the closed form
\begin{equation}
  \label{eq:ddpmpp-forward}
  \mathbf{x}_t=\sqrt{\bar{\alpha}_t}\mathbf{x}_0+\sqrt{1-\bar{\alpha}_t}\bar{\mathbf{z}}_t+\mathbf{Q}_t,
  \qquad
  \bar{\mathbf{z}}_t\sim\mathcal{N}(0,I),
  \qquad
  \mathbf{Q}_t=\left(\sum_{k=1}^{t}\sqrt{(1-\alpha_k)\prod_{i=k+1}^{t}\alpha_i}\right)\mathbf{q}.
\end{equation}
Here, $\mathbf{Q}_t$ accumulates the effect of the mean field $\mathbf{q}$ over
time and controls the expected drift of the process.

At inference, reverse diffusion starts from the input geometry
$\mathbf{x}'_T$ and iteratively generates the polycube structure $\mathbf{x}'_0$
by removing deformations predicted by a parameterized denoising network
$\mathbf{z}'_\theta(\mathbf{x}'_t,t,\mathbf{c})$. We set $\sigma_t^2=\beta_t$
and update samples by
\begin{equation}
  \label{eq:ddpmpp-reverse}
  \mathbf{x}'_{t-1}=
  \frac{1}{\sqrt{\alpha_t}}
  \left(
    \mathbf{x}'_t-\frac{\beta_t}{\sqrt{1-\bar{\alpha}_t}}
    \big(\mathbf{z}'_\theta(\mathbf{x}'_t,t,\mathbf{c})+\mathbf{q}'\big)
  \right)
  +\sigma_t\mathbf{z}'_t,
  \qquad
  \sigma_t^2=\beta_t,
  \qquad
  \mathbf{q}'=\frac{\sqrt{1-\bar{\alpha}_t}}{\sqrt{1-\alpha_t}}\mathbf{q}.
\end{equation}
We sample $\mathbf{z}'_t\sim\mathcal{N}(0,I)$ for $t>1$ and set $\mathbf{z}'_1=0$.

The denoising network $\mathbf{z}'_\theta(\cdot)$ follows a U-Net style design
with residual convolutional blocks, two downsampling stages, two upsampling
stages, and skip connections. The tensor
$\mathbf{x}'_t\in\mathbb{R}^{64\times 96\times 3}$ is treated as a three-channel
image-like representation whose channels correspond to the $X$, $Y$, and $Z$
coordinates. The timestep $t$ and the context $\mathbf{c}$ are embedded by fully
connected layers and fused with convolutional features through additive and
multiplicative modulation at multiple resolutions. Compared with the original
DDPM-Polycube setting on $32\times 32$, the main changes here are the higher
$64\times 96$ resolution and the 132-dimensional context encoding aligned with
the $G_{3\times 2\times 2}$ grid.

Training minimizes the mean squared error between the true noise and the
predicted noise:
\begin{equation}
  \label{eq:ddpmpp-loss}
  \mathcal{L}=\mathbb{E}_{\mathbf{x}_0,\mathbf{z},t}
  \left[
    \|\mathbf{z}-\mathbf{z}'_\theta(\mathbf{x}'_t,t,\mathbf{c})\|^2
  \right].
\end{equation}
The same cell-wise ordering, sorting, and reshaping rules described in
Section~\ref{sec:feature_extraction} are used in both training and inference to
keep the geometry tensor consistent with the context layout.

\section{Genus-guided context generation with hierarchical verification}
\label{sec:context-search-verification}
This section presents the inference method for generating the context vector
used in the context-conditioned reverse diffusion process. The proposed pipeline
supports two inference modes (Fig.~\ref{fig:Pipeline_Structure_V2}): a
\textbf{user-guided mode}, in which the context is fully or partially specified
by the user, and an \textbf{automated mode}, in which the context is generated
by the algorithm. In the original DDPM-Polycube
pipeline~\cite{yu_ddpm-polycube_2026}, automated inference is performed by
filtering candidate contexts according to the genus of the input triangular mesh
and then traversing the remaining candidates. This strategy is practical in the
$G_{2\times 1}$ setting, where the primitive set is limited and the number of
genus-consistent candidates remains small.  However, in the scalable setting
considered here, both the primitive set and the grid configuration are enlarged,
which causes the number of genus-consistent candidates to grow rapidly. As a
result, the original candidate-traversal strategy remains applicable in
principle, but its computational cost increases substantially because automated
inference would require many more reverse diffusion trials and candidate
verification steps.

To improve the scalability of inference, we introduce a new context generation
strategy under a unified context interface for both user-guided and automated
modes, and validate both intermediate and final results using a hierarchical
verification module. For clarity, this section focuses on the automated context
generation strategy. This emphasis does not exclude user guidance. On the
contrary, the pipeline also supports user-guided inference, and when the
provided guidance passes the genus-based consistency check and the hierarchical
verification module, inference is typically more efficient than in the fully
automated mode because it reduces the number of reverse diffusion trials and
candidate verification steps.

\subsection{Context generation}
Given an input geometry surface, the inference procedure first determines
whether user guidance is available. In the user-guided mode, the user may
provide either partial constraints, including per-geometry counts and spatial
distribution, or a complete 132-dimensional cell-wise one-hot encoded context
vector. In addition, when automated volumetric partitioning is not sufficiently
reliable, the user may specify volumetric splitting ratios to guide subregion
construction. If partial constraints are provided, they are converted into the
same internal 132-dimensional cell-wise one-hot context representation used
during training, so that the subsequent reverse diffusion process always
receives a context vector in a unified format.

After a context vector is obtained either directly from user input or by
converting partial constraints, we perform a genus-based consistency check. The
genus of the input geometry is computed from the Euler characteristic of its
triangular mesh. If the check fails, the user-provided context or constraints
are likely inconsistent with the topology of the input geometry. In this case,
the inference procedure asks the user to revise the specification or to switch
to automated context generation. If the check succeeds, the user-provided
context is used, and the resulting context-conditioned output is further
validated by the hierarchical verification module described in
Section~\ref{sec:hierarchical_verification}.

If no user context or constraints are provided, or if the user explicitly
requests automation, the pipeline invokes automated context generation. Instead
of traversing a large global candidate set directly, our method constructs the
global 132-dimensional cell-wise one-hot context by inferring locally consistent
labels from volumetric subregions. We first compute a temporary volumetric
representation of the input geometry by tetrahedralization. This temporary
tetrahedral volume is used only for context generation. We then partition it
into multiple subregions according to volumetric splitting ratios. The default
setting uses equal proportions, and user-defined ratios can be applied when
additional guidance is available.

For each subregion, which consists of a set of tetrahedral elements, we extract
its boundary surface as a triangular mesh and compute its genus. The subregion
genus provides a coarse topological constraint that restricts the candidate
primitive categories, because each primitive category has a fixed topological
type. This restriction is necessary but not sufficient. For example, when a
subregion has genus-0, multiple primitive categories remain feasible, so genus
alone cannot determine a unique local context. We therefore perform local
reverse diffusion inference to resolve this ambiguity.

For local inference, we use the extracted subregion boundary surface as the
input geometry. The surface is sampled into a point cloud tensor using the same
normalization and ordering rules as in Section~\ref{sec:feature_extraction}. We
then run context-conditioned reverse diffusion on this subregion input while
restricting candidate categories to the genus-feasible set implied by the
subregion genus. The resulting subregion candidate is then validated by the
hierarchical verification module. We first apply GOCC, as described later in
Algorithm~\ref{alg:stage1_gocc}, to perform a coarse occupancy and extent
validation under the current local context. If GOCC succeeds, it returns a
Boolean pass flag together with the verified active-cell set, denoted by
$\mathcal{A}$. At the subregion level, $\mathcal{A}$ contains the tuple for the
single active cell, including its local point subset and the decoded label
$c_j$. We then apply TCV, as described in Algorithm~\ref{alg:stage2_tcv}, only
to the active cells returned by GOCC in order to confirm local label-geometry
consistency by requiring the target template to be the nearest
competitor. Subregion candidates that pass both checks are accepted and
converted into verified local contexts. Finally, all accepted local contexts are
assembled into a global 132-dimensional cell-wise one-hot context vector
according to the fixed grid cell ordering.

Once the global context has been obtained, either from user guidance or from the
automated procedure, we run reverse diffusion on the full input geometry
conditioned on this context to generate a polycube structure candidate.  The
candidate is validated again by the same hierarchical verification module. If
verification succeeds, the polycube is accepted for subsequent parametric
mapping, all-hex control mesh generation, and volumetric spline construction. If
verification fails and the context originates from user input, the failure
indicates that the provided context is genus-consistent but still incompatible
with the learned deformation prior, and the inference procedure asks the user to
modify the specification or switch to automated context generation. If
verification fails and the context originates from automation, the failure
indicates that the temporary volumetric partition used for context construction
is not suitable under the default splitting strategy. In this case, the
inference procedure requests user guidance on partition ratios and then repeats
automated context generation.  When multiple verified contexts are available, we
select the one that yields the best hex-mesh quality, measured by the minimum
scaled Jacobian after mesh optimization in the physical domain.

\subsection{Hierarchical verification module}
\label{sec:hierarchical_verification}
The hierarchical verification module is used at two stages during
inference. During automated context generation, it validates subregion-level
candidates, where each subregion is treated as a single grid cell and the
associated context is a single 11-dimensional one-hot encoded context
vector. After a global context is assembled, the same module is applied again to
validate the polycube structure candidate generated from the input geometry,
where the grid contains 12 cells and the context is the 132-dimensional
cell-wise one-hot vector formed by concatenating 12 cell blocks.  In this way,
the same verification mechanism is shared by both the automated and user-guided
modes, as well as by both local and global inference stages.  This shared
verification interface improves reproducibility and eliminates the need for
manual inspection. The module is designed to satisfy two requirements. First, it
should discard clearly invalid candidates as early as possible in order to
reduce unnecessary reverse diffusion trials. Second, it should be sufficiently
selective to verify that each occupied cell agrees with the intended primitive
category, including its axis-aligned orientation variant. We meet these
requirements by combining a GOCC with a TCV driven by a penalty Chamfer distance
(PCD).

At the global level, GOCC takes as input the point cloud tensor
$P\in\mathbb{R}^{64\times96\times3}$ and the global cell-wise one-hot context
vector $L\in\{0,1\}^{132}$. The tensor $P$ is first reshaped into a point set in
$\mathbb{R}^3$, and is then partitioned according to the fixed unfolded
$4\times3$ block layout associated with the $G_{3\times2\times2}$ grid. The
placeholder points near the origin satisfying $\|p\|_2 \le 10^{-4}$ are
removed. These near-origin points correspond to the null-cell padding introduced
during tensor construction and should not be treated as valid geometry during
verification. For each cell $j\in\{1,\dots,12\}$, GOCC decodes the target state
$c_j\in\{0,\dots,10\}$ from the corresponding 11-dimensional one-hot block in
$L$, where $c_j=0$ denotes the null state and $c_j>0$ denotes the label of one
of the primitive categories. Let $P_j$ denote the local point subset assigned to
cell $j$. GOCC then checks the decoded label against two coarse geometric
indicators: the point count $|P_j|$ and the spatial extent $S_j$, where $S_j$ is
defined as the maximum axis-wise span of the axis-aligned bounding box of
$P_j$. In our implementation, occupied cells are required to contain at least 10
points and to satisfy the activity threshold $S_j \ge \tau_{\text{active}}$,
while null cells must contain fewer than 10 points. Here, the threshold value 10
serves as a coarse occupancy discriminator, and $\tau_{\text{active}}$ controls
the minimum geometric extent required for an occupied cell to be considered
active. This component serves as a fast coarse filter that rejects candidates
when the predicted point distribution across cells does not match the occupancy
labels encoded in the context. If GOCC succeeds, it returns the verified
active-cell set $\mathcal{A}=\{(j,P_j,c_j)\}$, where each tuple contains the
cell index, the corresponding local point subset, and the decoded target
label. The returned set $\mathcal{A}$ allows TCV to operate directly on the
active cells identified by GOCC. Algorithm~\ref{alg:stage1_gocc} gives the
global-level version of GOCC for the 12-cell layout. The subregion-level case is
analogous, with a single local cell and the corresponding 11-dimensional one-hot
label.

Candidates that pass GOCC are then tested by TCV to confirm local label-geometry
consistency. Specifically, TCV is applied only to the active cells returned by
GOCC rather than to all 12 cells. For each active cell, we center the local
subset $P_j$ by subtracting its centroid $\bar{p}=\mathrm{mean}(P_j)$ and obtain
the centered subset $P'_j=P_j-\bar{p}$. We perform the same centering operation
for every template $T_k$ in the template library $\mathbb{T}$, yielding centered
templates $T'_k=T_k-\mathrm{mean}(T_k)$. Here, $\mathbb{T}$ is the primitive
library shown in Fig.~\ref{fig:grid_unfold}(a), which contains ten axis-aligned
categories derived from three base primitives and is used both for dataset
generation and for template matching in TCV. We then evaluate the PCD between
$P'_j$ and all centered templates. Let
$d_{\text{target}}=\mathrm{PCD}(P'_j,T'_{c_j},\tau_{\text{CD}},p)$ denote the
distance to the target template specified by the decoded label $c_j$, and let
$k^*=\arg\min_k \mathrm{PCD}(P'_j,T'_k,\tau_{\text{CD}},p)$ denote the index of
the nearest competing template. The intended label is accepted only when two
conditions are satisfied. First, the target template must be sufficiently close,
with $d_{\text{target}} \le \tau_{\text{CD}}$. Second, it must also be the
nearest template among all competitors, which requires $k^* = c_j$. Here,
$\tau_{\text{CD}}$ is the acceptance threshold of the distance metric, while $p$
controls the penalty strength applied to large deviations.  This
nearest-template matching criterion confirms both the primitive category and the
axis-aligned orientation variant encoded in the context.
Algorithm~\ref{alg:stage2_tcv} summarizes the single-cell TCV procedure, which
is applied to each active cell in both subregion-level and global-level
verification.

The PCD is defined as
\begin{equation}
  \label{eq:penalty_cd}
  \mathrm{PCD}(X, Y, \tau_{\text{CD}}, p) = \frac{1}{2} \left(
    \mathbb{E}_{x \in X} \left[f\!\left(\min_{y \in Y} \|x-y\|^2\right)\right] +
    \mathbb{E}_{y \in Y} \left[f\!\left(\min_{x \in X} \|x-y\|^2\right)\right]
  \right),
\end{equation}
where the penalty function is
\begin{equation}
  \label{eq:penalty_func}
  f(d) =
  \begin{cases}
    d, & d \leq \tau_{\text{CD}},\\
    d\left(1+\left(\frac{d}{\tau_{\text{CD}}}\right)^p\right), & d > \tau_{\text{CD}}.
  \end{cases}
\end{equation}
Here, $\tau_{\text{CD}}$ defines the threshold beyond which nonlinear penalty
amplification is applied, and $p$ controls the amplification strength once the
distance exceeds this threshold. By amplifying large deviations, this metric
becomes more sensitive to structural mismatches that would otherwise be
under-penalized by the standard Chamfer distance.

\begin{algorithm}[htbp]
  \caption{Grid Occupancy Consistency Check (GOCC, global-level version)}
  \label{alg:stage1_gocc}
  \begin{algorithmic}[1]
    \Require Point cloud tensor $P \in \mathbb{R}^{64 \times 96 \times 3}$,
    global cell-wise one-hot context vector $L \in \{0,1\}^{132}$, activity threshold $\tau_{\text{active}}$
    \Ensure Pass/fail flag and the verified active-cell set $\mathcal{A}$
    \State Reshape $P$ into a point set in $\mathbb{R}^3$
    \State Partition the reshaped point set into 12 local subsets according to
    the fixed unfolded $4\times3$ block layout
    \State Remove placeholder points satisfying $\|p\|_2 \leq 10^{-4}$
    \State $\mathcal{A} \leftarrow \emptyset$
    \For{$j = 1$ \textbf{to} $12$}
      \State $c_j \leftarrow \arg\max\!\big(L[(j-1)\times 11 + 1 : j\times 11]\big)$
      \State $P_j \leftarrow$ the local point subset associated with cell $j$
      \If{$c_j > 0$}
        \If{$|P_j| < 10$}
          \State \Return \textbf{False}, $\emptyset$
        \EndIf
        \State $S_j \leftarrow \max_{\text{axis}}\!\big(\max(P_j) - \min(P_j)\big)$
        \If{$S_j < \tau_{\text{active}}$}
          \State \Return \textbf{False}, $\emptyset$
        \EndIf
        \State $\mathcal{A} \leftarrow \mathcal{A} \cup \{(j, P_j, c_j)\}$
      \Else
        \If{$|P_j| \geq 10$}
          \State \Return \textbf{False}, $\emptyset$
        \EndIf
      \EndIf
    \EndFor
    \State \Return \textbf{True}, $\mathcal{A}$
\end{algorithmic}
\end{algorithm}

\begin{algorithm}[htbp]
  \caption{Template Competition Verification (TCV, single-cell version)}
  \label{alg:stage2_tcv}
  \begin{algorithmic}[1]
    \Require Local subset $P_j$, target label $c_j$, template library $\mathbb{T}$, distance threshold $\tau_{\text{CD}}$, penalty power $p$
    \Ensure Pass/fail flag and the target similarity score $d_{\text{target}}$
    \State $\bar{p} \leftarrow \mathrm{mean}(P_j)$
    \State $P'_j \leftarrow P_j - \bar{p}$
    \ForAll{$T_k \in \mathbb{T}$}
      \State $T'_k \leftarrow T_k - \mathrm{mean}(T_k)$
    \EndFor
    \State $d_{\text{target}} \leftarrow \mathrm{PCD}(P'_j, T'_{c_j}, \tau_{\text{CD}}, p)$
    \State $k^* \leftarrow \arg\min_k \mathrm{PCD}(P'_j, T'_k, \tau_{\text{CD}}, p)$
    \If{$(k^* = c_j) \land (d_{\text{target}} \leq \tau_{\text{CD}})$}
      \State \Return \textbf{True}, $d_{\text{target}}$
    \Else
      \State \Return \textbf{False}, $d_{\text{target}}$
    \EndIf
  \end{algorithmic}
\end{algorithm}

\section{Results and discussion}
\label{sec:results}
This section evaluates the proposed SDDPM pipeline from three
aspects. First, we study the training behavior of the diffusion model under the
expanded primitive set and the 3D grid configuration.  Second, we evaluate
inference robustness in both the user-guided mode, where the context is provided
as a complete 132-dimensional vector or derived from partial constraints, and
the automated mode, where the context is generated by genus-guided local
inference together with hierarchical verification. Third, we assess performance
in all-hex control mesh generation and volumetric spline construction. We report the
polycube generation time, including reverse diffusion and verification, and we
report mesh quality after all-hex control mesh optimization. Mesh quality is quantified by
the minimum scaled Jacobian. All experiments were conducted on a workstation
equipped with dual Intel Xeon Platinum 8458P CPUs (88 cores in total), 256 GB of
RAM, and an NVIDIA L20 GPU with 48 GB of VRAM, operating under CUDA 12.4.

\subsection{Training performance on expanded primitives and $G_{3\times 2\times 2}$}
We train SDDPM using the dataset described in
Sections~\ref{sec:grid_config}--\ref{sec:feature_extraction}. Compared with
DDPM-Polycube, this work extends the primitive set from two primitive geometries
to three by introducing the BHC, together with axis-oriented variants. It also
moves beyond the 1D $G_{2\times 1}$ setting and adopts the 3D
$G_{3\times 2\times 2}$ grid configuration with 12 cells, which results in a
132-dimensional cell-wise one-hot context vector. In addition, the input tensor
resolution is increased to $64\times 96\times 3$. The training objective
minimizes the mean squared error between the ground-truth noise and the
predicted noise at each timestep (Eq.~\eqref{eq:ddpmpp-loss}). In the forward
diffusion, we use a nonzero-mean Gaussian noise model
$\mathcal{N}(\mathbf{q},I)$ so that the endpoint of the forward process remains
a geometry-like distribution rather than collapsing to a standard Gaussian. This
design follows the DDPM-Polycube formulation~\cite{yu_ddpm-polycube_2026} and is
important for stable polycube generation from the input geometry during reverse
diffusion.

\begin{figure}[pos=htbp]
  \centering
  \begin{tikzpicture}
    \node[anchor=south west,inner sep=0] (image) at (0,0) {\includegraphics[width=\linewidth]{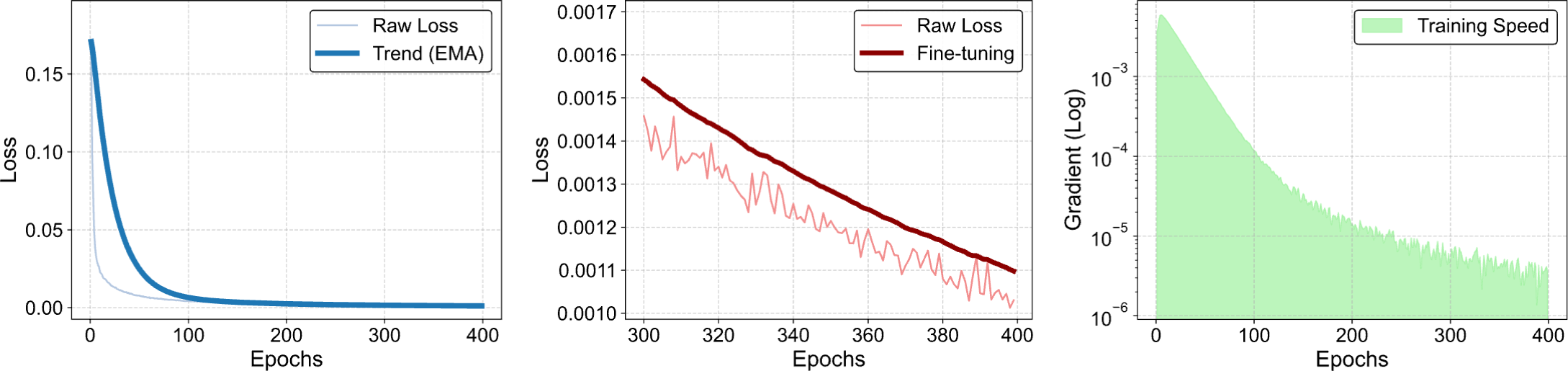}};
    \begin{scope}[x={(image.south east)},y={(image.north west)}]
       \node at (0.17,-0.05) {\small (a)};
       \node at (0.52,-0.05) {\small (b)};
       \node at (0.88,-0.05) {\small (c)};
    \end{scope}
  \end{tikzpicture}
  \caption{Training behavior of SDDPM on the expanded primitive
    set and the $G_{3\times 2\times 2}$ grid. (a) Overall training loss over 500
    epochs. (b) Enlarged view of the final 100 epochs. (c) Gradient of the
    training loss.}
  \label{fig:training_loss}
\end{figure}
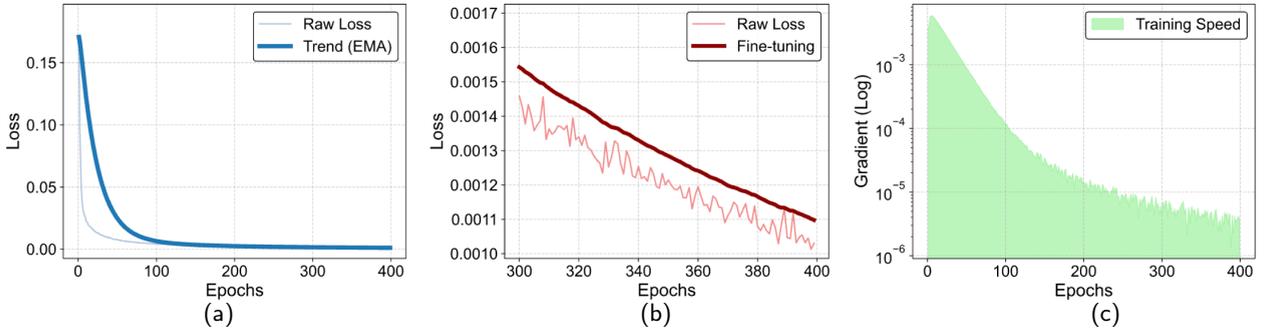

The diffusion process is configured with 500 timesteps. The model is trained
with a batch size of 128 over 500 epochs. The initial learning rate is set to
$1\times 10^{-4}$ and is adjusted using a linear decay strategy,
$\eta_k = \eta_{0} \times \left(1 - \frac{k}{K}\right)$, where $\eta_0$ is the
initial learning rate, $k$ is the current epoch, and $K$ is the total number of
epochs. The Adam optimizer is used throughout training. The training takes
approximately 35 hours on the workstation described
above. Fig.~\ref{fig:training_loss} summarizes the training behavior of the
model. Fig.~\ref{fig:training_loss}(a) shows the overall decrease in the
training loss over all epochs, indicating steady
convergence. Fig.~\ref{fig:training_loss}(b) zooms into the final 100 epochs and
shows that the loss continues to decrease smoothly during the late stage of
training, which suggests stable fine-tuning. Fig.~\ref{fig:training_loss}(c)
plots the gradient of the loss and provides an additional view of the learning
efficiency during optimization.

\subsection{Inference with user-guided contexts and partial constraints}
\label{sec:user_guided_results}
We first evaluate the pipeline under user guidance. The user can provide either
a complete 132-dimensional cell-wise one-hot context vector or partial constraints,
including per-geometry counts and spatial distribution. These partial
constraints are converted to the same internal context format used during training
(Section~\ref{sec:context-search-verification}). After a context is obtained, we
compute the genus of the input geometry and perform a genus-based consistency
check before running reverse diffusion.

\begin{figure}[pos=htbp]
  \centering
  \includegraphics[width=0.8\linewidth]{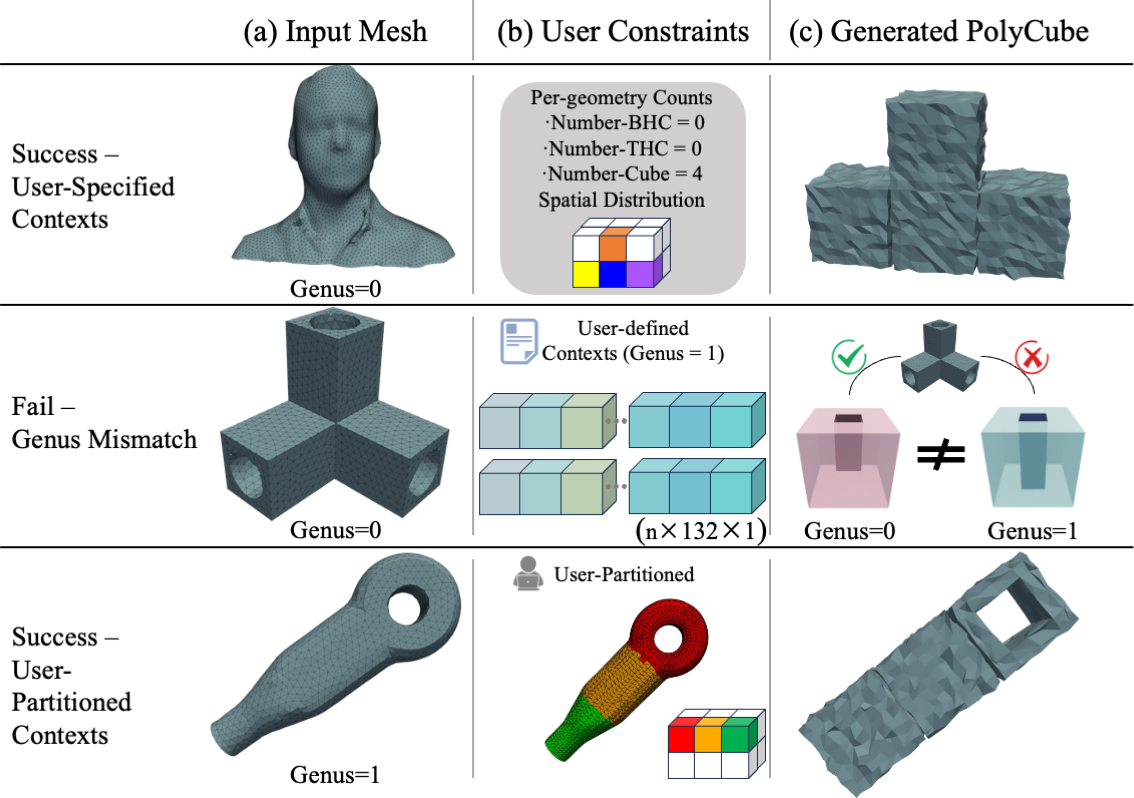}
  \caption{Representative examples of user-guided inference in SDDPM. The first
    row shows generation under high-level user constraints. The second row shows
    rejection of user-provided contexts that are inconsistent with the input
    genus. The third row shows generation under user-specified volumetric
    partition guidance.}
  \label{fig:user_guided_examples}
\end{figure}

Our experiments (Fig.~\ref{fig:user_guided_examples}) illustrate the
effectiveness of the proposed inference procedure across representative
user-guided cases. In the first row, the user specifies only high-level
constraints, such as the per-geometry counts and spatial distribution. The model
then generates a valid polycube structure that satisfies the prescribed
requirements. In the second row, the user provides a context whose genus is
inconsistent with that of the input geometry. The genus-based consistency check
detects the topological mismatch and blocks the diffusion process, thereby
avoiding unnecessary computation. In the third row, the user provides guidance
on subregion partitioning in the form of user-defined volumetric splitting
ratios. The generated polycube structure is consistent with the input
geometry. Compared with the fully automated mode, user-guided inference can be
more efficient when the provided guidance passes the genus-based consistency
check and the subsequent hierarchical verification. This is because the search
space is reduced and fewer reverse diffusion trials and candidate verification
steps are required.

\subsection{Inference with automated context generation}
\label{sec:auto_context_results}
We next evaluate the fully automated mode, which is invoked when the user does
not provide contexts or constraints, or explicitly chooses automation. The goal
is to construct a global 132-dimensional context without traversing a large
global candidate space. Following Section~\ref{sec:context-search-verification},
we tetrahedralize the input shape temporarily and partition the volume into
subregions. For each subregion, we extract its boundary surface, compute its
genus, restrict the local candidate primitive categories accordingly, and run a
local reverse diffusion process to generate a local candidate geometry, from
which a candidate label is verified. Each subregion candidate is then validated
immediately by the hierarchical verification module. In particular, it is first
checked by GOCC and then verified by TCV. Verified subregion labels are
aggregated into a global 132-dimensional context aligned with the grid
layout. Finally, reverse diffusion is run on the full geometry using the
assembled global context, followed by the same verification procedure.

\begin{figure}[pos=htbp]
  \centering
  \begin{tikzpicture}
    \node[anchor=south west,inner sep=0] (image) at (0,0) {\includegraphics[width=\linewidth]{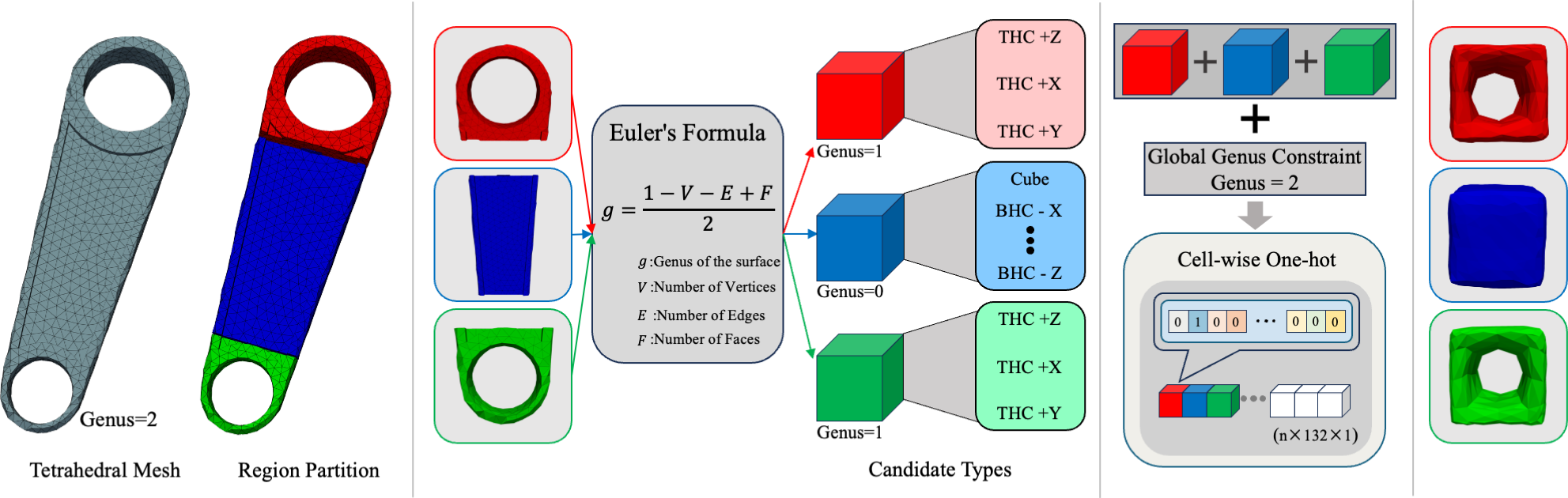}};
    \begin{scope}[x={(image.south east)},y={(image.north west)}]
      \node at (0.12,-0.05) {\small (a)};
      \node at (0.47,-0.05) {\small (b)};
      \node at (0.8,-0.05) {\small (c)};
      \node at (0.96,-0.05) {\small (d)};
    \end{scope}
  \end{tikzpicture}
  \caption{Automated context generation and global assembly in the proposed
    inference pipeline. (a) Temporary tetrahedralization of the input geometry
    and partition of the volume into subregions. (b) Computation of the genus
    for each subregion and restriction of feasible primitive categories. (c)
    Aggregation of verified local labels into a global 132-dimensional cell-wise one-hot
    context vector aligned with the $G_{3\times 2\times 2}$ grid. (d) Final
    polycube structure generated by reverse diffusion conditioned on the
    assembled global context.}
  \label{fig:auto_context_examples}
\end{figure}
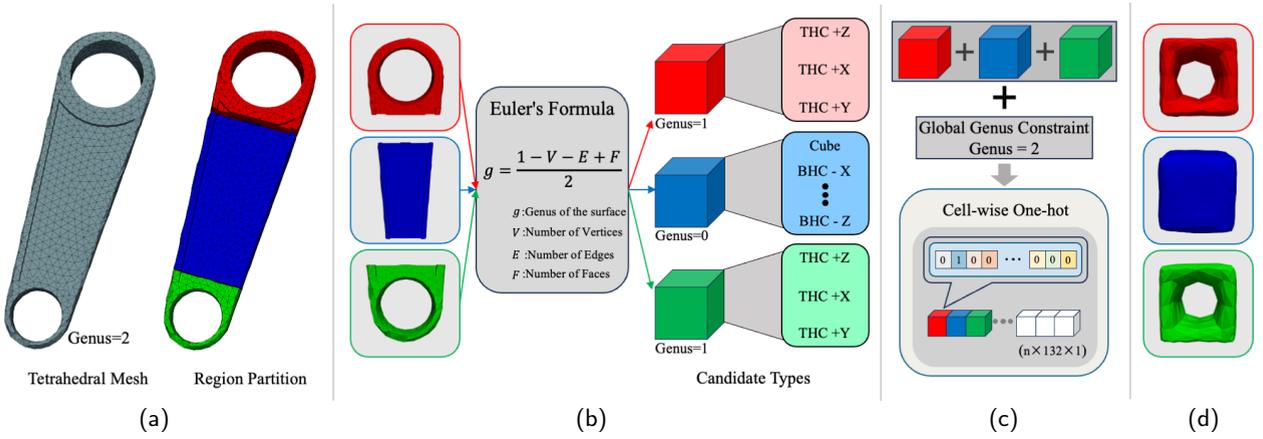

Fig.~\ref{fig:auto_context_examples} illustrates this automated context
generation process. These results demonstrate that the proposed automated
context generation strategy provides a practical way to reduce the inference
cost associated with large candidate spaces. By combining local genus
constraints with GOCC and TCV, the method avoids direct traversal of the full
global context space while still producing valid and topology-consistent
polycube structures. This automated procedure is particularly useful for complex
geometries when reliable user guidance is unavailable or difficult to provide.

\subsection{Effectiveness of hierarchical verification}
We further evaluate the effectiveness of the hierarchical verification module
introduced in Section~\ref{sec:hierarchical_verification}. The module consists
of two complementary stages: GOCC and TCV. GOCC evaluates whether the predicted
point distribution agrees with the context in terms of cell-wise occupancy and
coarse geometric extent. TCV then evaluates whether each occupied cell matches
the intended primitive category by requiring the target template to be the
nearest competitor under the PCD. Together, these two
stages provide a coarse-to-fine verification mechanism for both local subregion
inference and final global context validation.

\begin{figure}[pos=htbp]
  \centering
  \begin{tikzpicture}
    \node[anchor=south west,inner sep=0] (image) at (0,0) {\includegraphics[width=\linewidth]{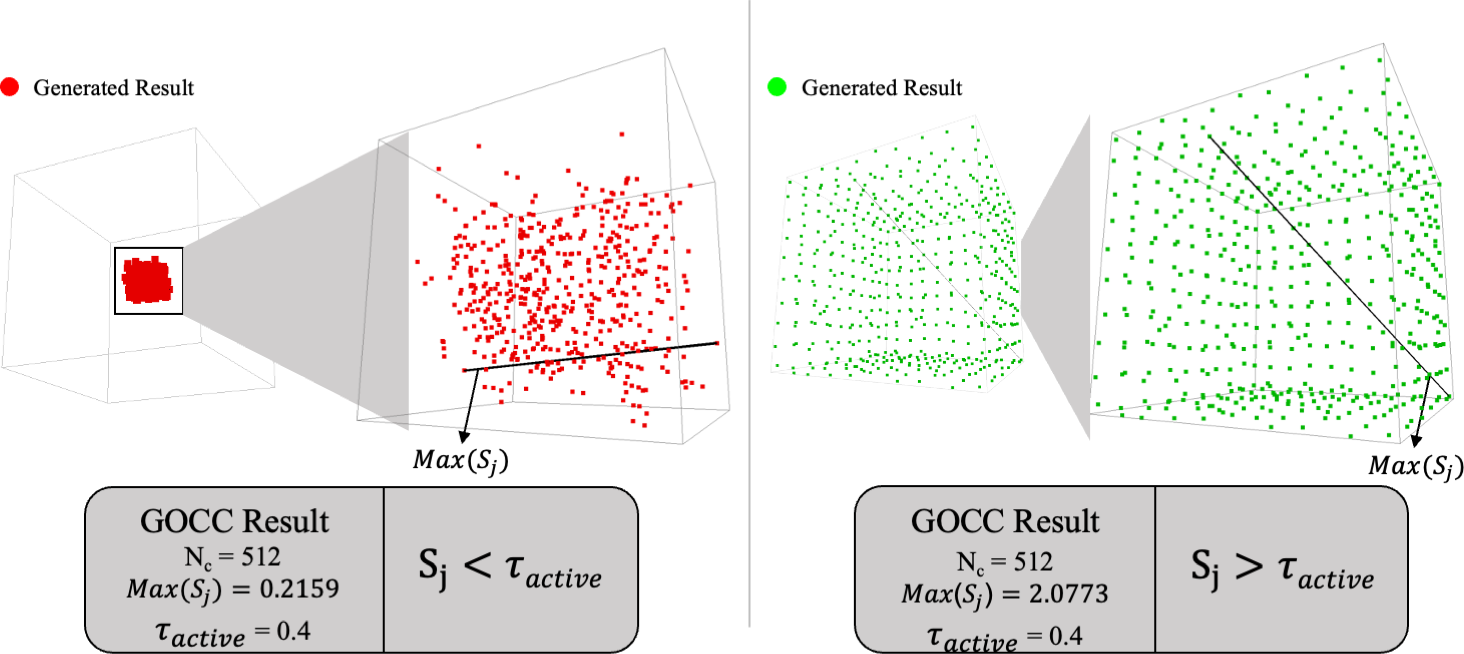}};
    \begin{scope}[x={(image.south east)},y={(image.north west)}]
      \node at (0.24,-0.05) {\small (a)};
      \node at (0.76,-0.05) {\small (b)};
    \end{scope}
  \end{tikzpicture}
  \caption{Examples of GOCC. (a) Failure case, where the generated local point
    subset does not exhibit sufficient occupancy in the target cell. (b) Success
    case, where the generated point distribution exhibits sufficient occupancy
    in the target cell.}
  \label{fig:gocc_examples}
\end{figure}

To illustrate the behavior of the verification module at the local level,
Fig.~\ref{fig:gocc_examples} presents representative examples of GOCC under
unsuitable and suitable contexts. In Fig.~\ref{fig:gocc_examples}(a), an
unsuitable context leads to a GOCC failure. This failure is typically
characterized by an insufficient local point count or by a spatial extent below
the activation threshold, indicating that the predicted point distribution does
not match the occupancy encoded in the context. In contrast,
Fig.~\ref{fig:gocc_examples}(b) shows a case in which the occupied cell exhibits
sufficient point support and extent, so that the candidate passes GOCC. These
examples demonstrate that GOCC serves as an effective coarse filter for
rejecting candidates with occupancy or extent inconsistencies at low cost.

\begin{figure}[pos=htbp]
  \centering
    \begin{tikzpicture}
      \node[anchor=south west,inner sep=0] (image) at (0,0) {\includegraphics[width=\linewidth]{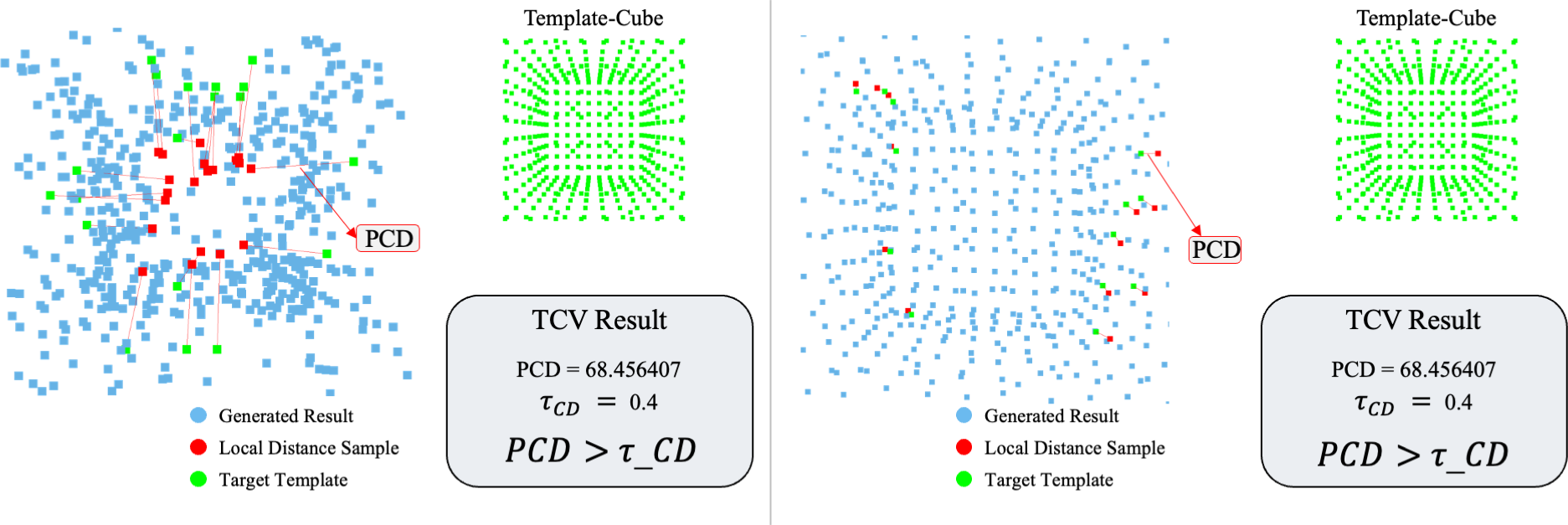}};
    \begin{scope}[x={(image.south east)},y={(image.north west)}]
      \node at (0.24,-0.05) {\small (a)};
      \node at (0.76,-0.05) {\small (b)};
    \end{scope}
  \end{tikzpicture}
  \caption{Examples of TCV. (a) Failure case, where the occupied local geometry
    does not match the target primitive category. (b) Success case, where the
    occupied local geometry matches the target primitive category, with the
    target template being the nearest competitor and the PCD remaining below the
    acceptance threshold.}
  \label{fig:tcv_examples}
\end{figure}

After a candidate passes GOCC, it is further examined by
TCV. Fig.~\ref{fig:tcv_examples}(a) shows a case in which the local subset is
occupied but the generated geometry still deviates from the intended
template. In this case, the PCD to the target template is large, or the nearest
competitor does not match the intended category, and the candidate is therefore
rejected by TCV. In contrast, Fig.~\ref{fig:tcv_examples}(b) shows a successful
case in which the target template is also the nearest competitor and the
corresponding PCD remains below the acceptance threshold. These examples
indicate that TCV provides a finer level of verification by enforcing
consistency between the generated local geometry and the target primitive
category, including its orientation variant.

To quantitatively assess the efficiency of this hierarchical verification
strategy, we conduct a stage-wise analysis on the Triple-Hole Vertical Ring
Bracket model shown in Fig.~\ref{fig:Pipeline_Structure_V2}. The theoretical
global context space consists of \(11^{12}\) possible assignments, since each of
the 12 cells can take one of 11 states. Because only four designated cells are
activated in this example and the remaining eight cells are fixed to the null
state in this example, the candidate space is immediately reduced to \(11^4\).
Applying the genus constraint of the input geometry (\(g=3\)) further reduces
the number of feasible candidate contexts to 189. These 189 candidates are then
processed by local verification, which consists of local GOCC and local TCV,
reducing the candidate set to 36. At the global verification stage, GOCC
achieves a pass rate of 27.78\% with respect to these 36 locally verified
candidates. The candidates that pass global GOCC are then evaluated by global
TCV, which achieves a pass rate of 60\% relative to the GOCC-passed subset. As a
result, the final verified candidate set contains 6 contexts.

These results demonstrate that the proposed multi-stage verification strategy
substantially reduces the effective search space while maintaining strict
topological and geometric consistency. Local verification eliminates a large
portion of infeasible contexts before full global verification is required, and
the subsequent global GOCC and TCV further remove candidates that are
inconsistent with the intended occupancy pattern and primitive
category. Although the verification procedure is conservative, especially at the
TCV stage, it provides a robust and label-consistent basis for subsequent
polycube generation, all-hex control mesh generation, and volumetric spline
construction. Overall, the combined use of local verification and global
hierarchical verification improves the reliability of both automated and
user-guided inference and substantially reduces the need for manual inspection
in the later stages of the pipeline.


\subsection{Performance on geometries of different genus}
In this section, we evaluate the performance of SDDPM on models
with varying genus levels. The results demonstrate its ability to generate
polycube structures on the $G_{3\times 2\times 2}$ grid, to handle more diverse
local features introduced by the expanded primitive set, and to produce valid
polycube assemblies under automated contexts.

\begin{figure}[pos=htbp]
  \centering
  \begin{tabular}{c}
    \includegraphics[width=\linewidth]{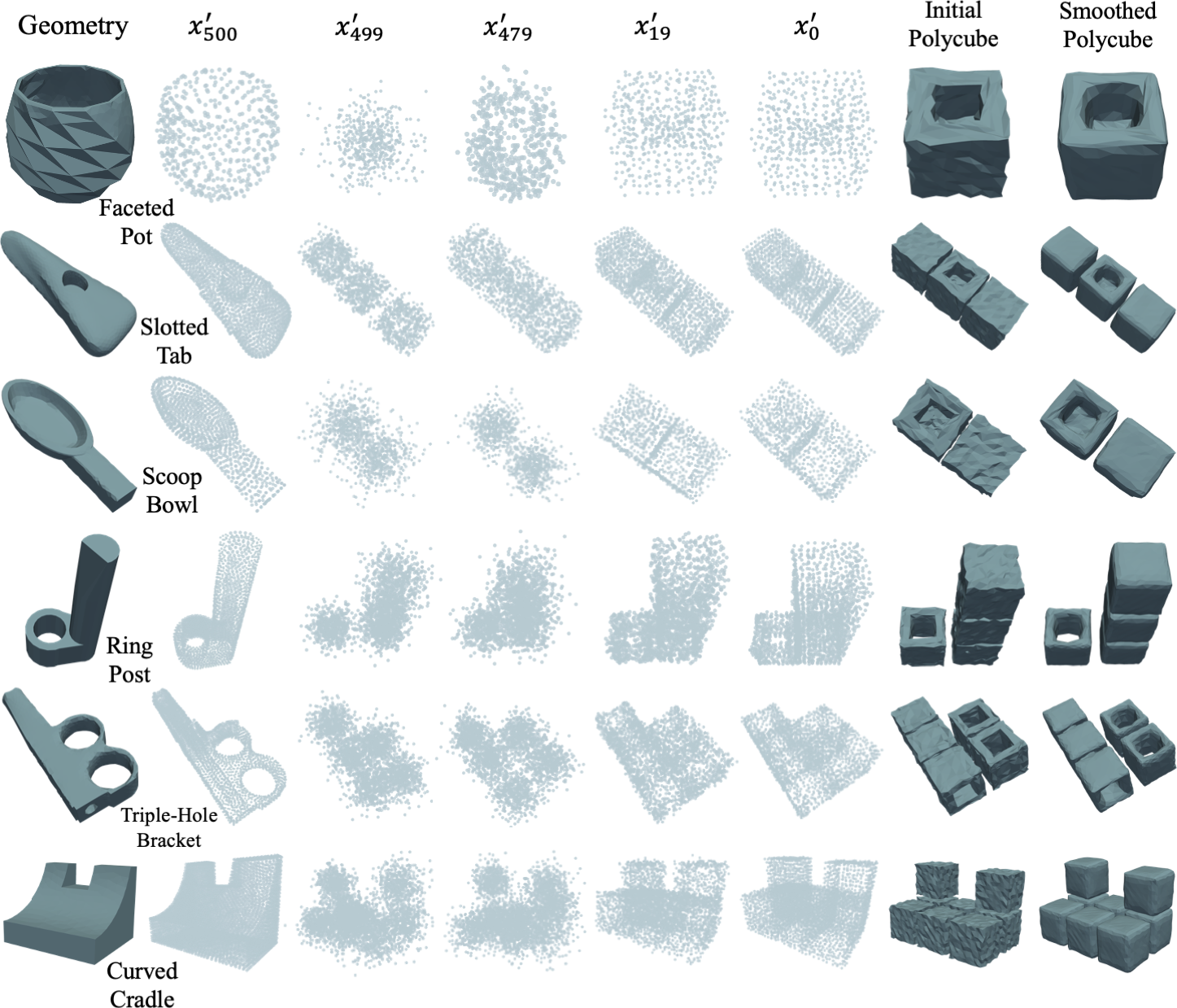}
  \end{tabular}
  \caption{Visualization of the reverse diffusion process for polycube
    generation. From left to right, the model progressively removes deformation
    from the input geometry at $t=500$ and recovers a topology-consistent
    polycube representation at $t=0$.}
  \label{fig:reverse_diffusion}
\end{figure}

Fig.~\ref{fig:reverse_diffusion} visualizes the reverse diffusion process from
$t=500$ to $t=0$. At $t=500$, the representation corresponds to the input
geometry, which can be interpreted as a polycube structure with accumulated
small-scale deformations. As the denoising process proceeds, the model
progressively removes these deformations and organizes the point cloud into a
polycube structure under the learned deformation prior conditioned on the
132-dimensional context. This figure illustrates how the reverse diffusion
process gradually recovers the target polycube structure from the input
geometry.

We first consider genus-0 geometries. These models include both simple cube-like
objects and models containing hole-like features that do not change the global
genus. With the added BHC primitive and its orientation variants,
the pipeline can explicitly represent these local features and generate polycube
structures that are consistent with the input geometry. In particular, the
method uses BHC variants to represent blind-hole regions while preserving the
genus-0 structure of the overall assembly.

We then consider genus-1 geometries. For models with one through-hole, the
pipeline generates polycubes that contain at least one THC cell and assigns its
axis orientation according to the dominant direction of the hole feature. The
results show that the method can place and orient THC variants consistently
across different shapes and deformations, and can combine THC cells with cube or
BHC cells when additional local features exist.

\begin{figure}[pos=htbp]
  \centering
  \begin{tabular}{c}
    \includegraphics[width=0.9\linewidth]{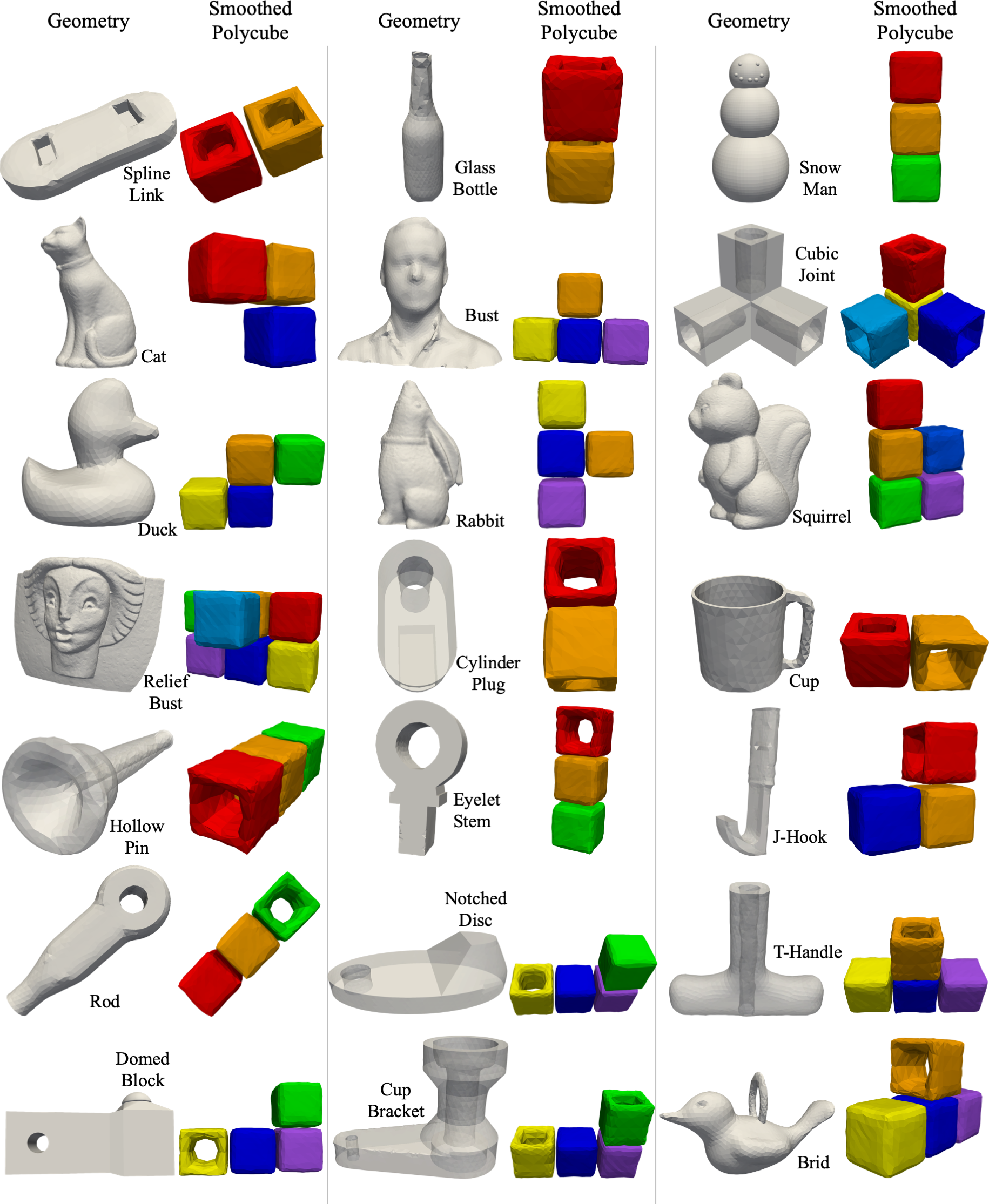}
  \end{tabular}
  \caption{Representative polycube generation results on geometries of different
    genus (Part~I). }
  \label{fig:partial_results_part1}
\end{figure}

\begin{figure}[pos=htbp]
  \centering
  \begin{tabular}{c}
    \includegraphics[width=\linewidth]{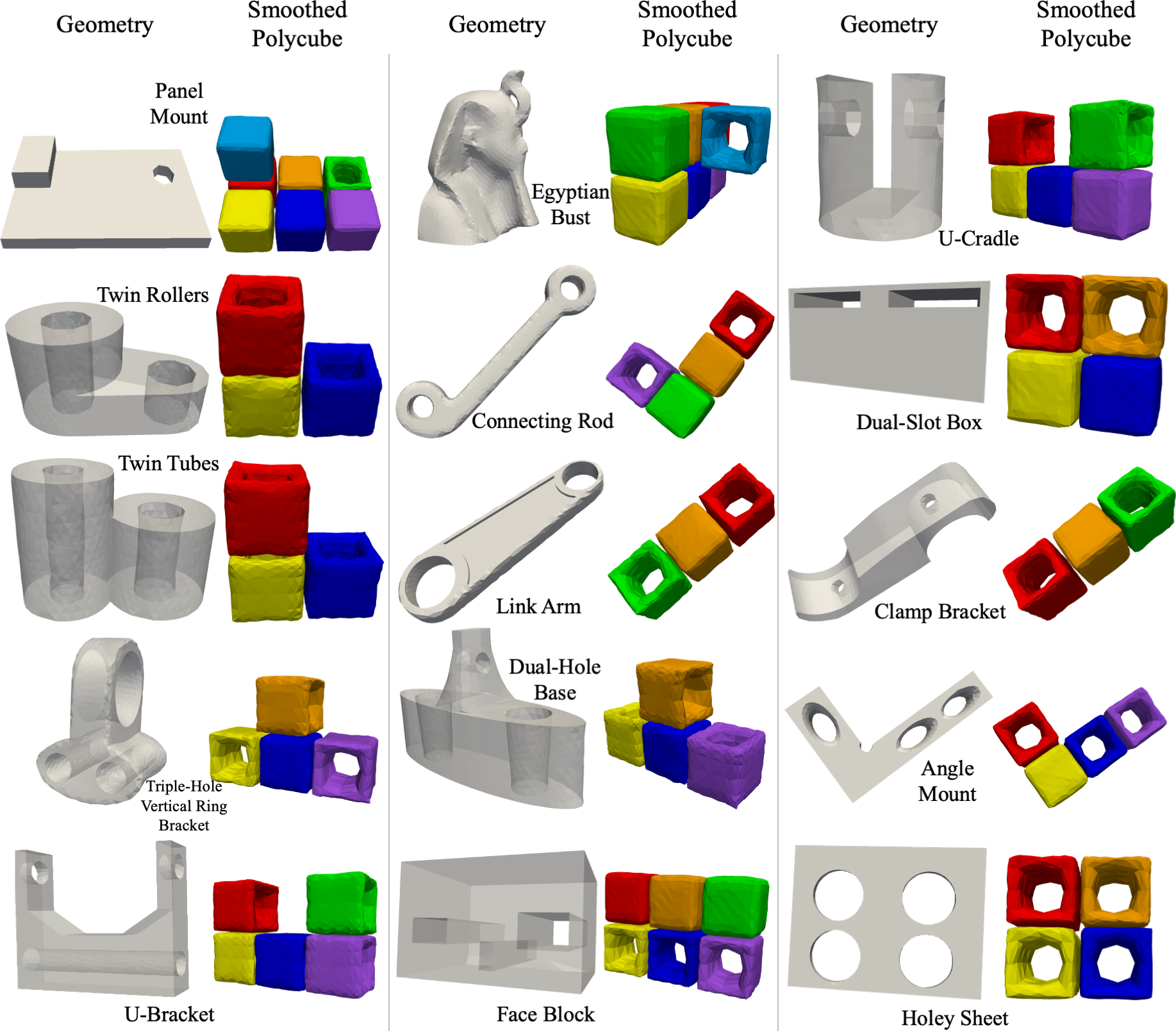}
  \end{tabular}
  \caption{Representative polycube generation results on geometries of different
    genus (Part~II). }
  \label{fig:partial_results_part2}
\end{figure}

Representative results for genus-0 and genus-1 models are shown in
Fig.~\ref{fig:partial_results_part1}. These examples include low-genus
geometries as well as models at the transition to more complex topologies. The
results demonstrate that the expanded primitive set allows the pipeline to
handle both blind-hole and through-hole structures in a unified
manner. Additional representative results for models with genus 2 and higher are
shown in Fig.~\ref{fig:partial_results_part2}. These cases require the pipeline
to capture multiple independent through-hole features and to assemble multiple
occupied cells into topology-consistent polycube structures on the 12-cell grid.

Across these examples, SDDPM generates polycube structures that remain
consistent with the topology of the input geometry. Compared with the previous
$G_{2\times 1}$ setting~\cite{yu_ddpm-polycube_2026}, the fixed
$G_{3\times 2\times 2}$ configuration provides more degrees of freedom, which
reduces the need to force complex geometries into a small number of cells and
supports the generation of more stable polycube structures under larger
geometric deformations. Overall, the experimental results in
Figs.~\ref{fig:reverse_diffusion}--\ref{fig:partial_results_part2} demonstrate
that the expanded primitive set and the higher-dimensional grid improve the
representational capacity of the pipeline. Together with hierarchical
verification, the pipeline reliably produces topology- and label-consistent
polycube structures across geometries of different genus, thereby providing a
robust foundation for subsequent all-hex control mesh generation.

We also note that, except for the Faceted Pot model, the generated polycube
structures of the remaining test cases do not have exact matches in the training
set under the cell-wise one-hot context representation on the
$G_{3\times 2\times 2}$ grid. These cases can therefore be regarded as unseen
polycube structure configurations, since their complete cell-wise one-hot
configurations are not explicitly present in the training data. This observation
suggests that SDDPM can generate valid polycube structures beyond exact training
configurations. This behavior is consistent with the deformation-based strategy
of the proposed method, which learns deformation priors from the primitive set
and grid-based compositions, and applies them to new test geometries during
inference. Unlike the template-mapping strategy used in
DL-Polycube~\cite{yu_dl-polycube_2025}, this strategy does not depend on
predefined polycube templates for such cases.


\subsection{Ablation and scalability analysis}
\label{sec:ablation_scalability}

In this subsection, we evaluate the effect of the introduced BHC primitive
through an ablation study, and analyze the scalability of the automated
inference procedure under different levels of structural complexity. To evaluate
the role of the BHC primitive, we conduct a controlled ablation experiment on
the Spline Link model shown in Fig.~\ref{fig:partial_results_part1} (the first
model in the first row). This model is a representative genus-0 case containing
blind-hole features. In this experiment, the input geometry, the trained
diffusion model, and the inference procedure are kept unchanged, while only the
primitive category label of the corresponding cell is modified at inference
time.  Specifically, we compare the proposed setting, in which the local feature
is assigned the \texttt{BHC+Z} primitive category label, with an ablation
setting in which the same feature is forced to use the \texttt{Cube} label.

\begin{figure}[pos=htbp]
  \centering
  \begin{tikzpicture}
    \node[anchor=south west,inner sep=0] (image) at (0,0) {\includegraphics[width=\linewidth]{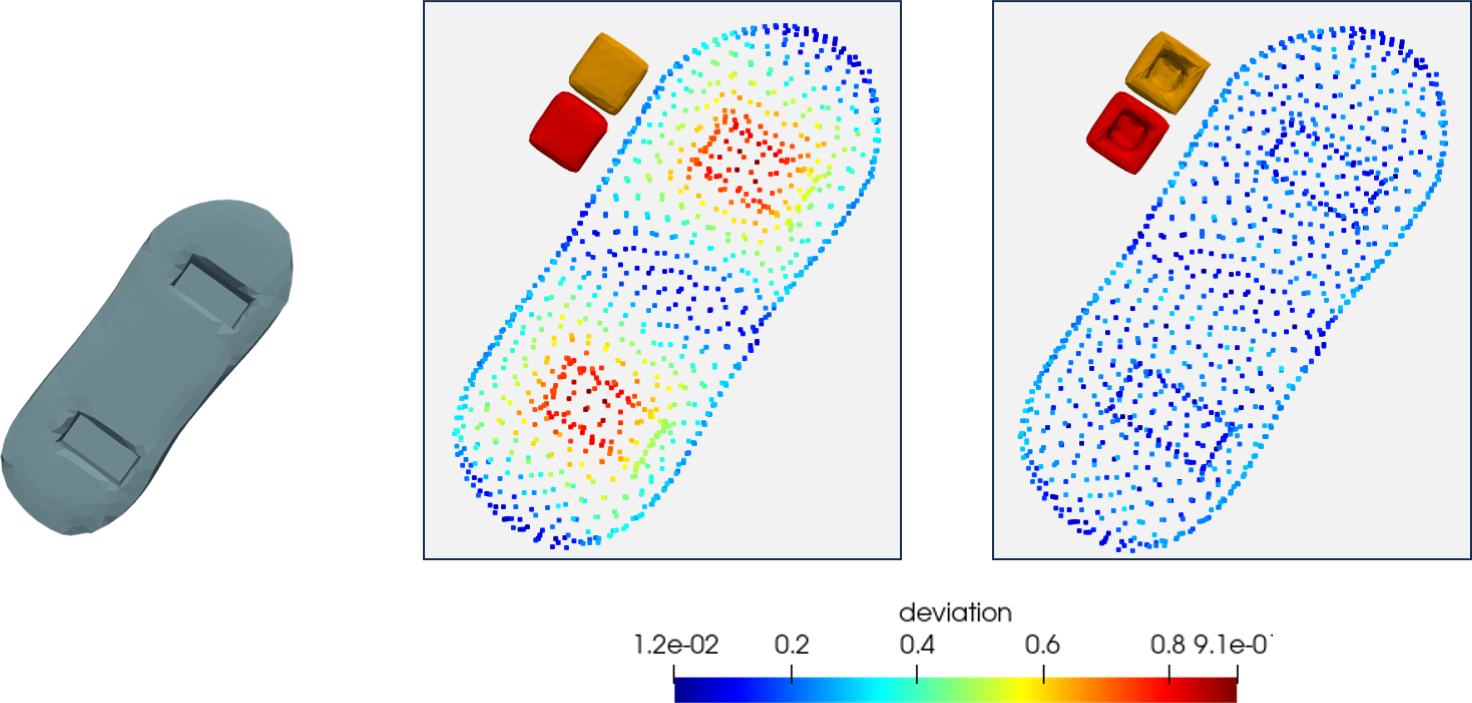}};
    \begin{scope}[x={(image.south east)},y={(image.north west)}]
      \node at (0.12,-0.05) {\small (a)};
      \node at (0.44,-0.05) {\small (b)};
      \node at (0.84,-0.05) {\small (c)};
    \end{scope}
  \end{tikzpicture}
  \caption{Ablation study on the BHC primitive. (a) Input geometry with
    blind-hole features. (b) Result without a BHC primitive label, where the
    generated polycube fails to preserve the intended local blind-hole feature
    and exhibits large geometric deviations near the blind-hole region.  (c)
    Result with the \texttt{BHC+Z} primitive label, where the blind-hole feature
    is better preserved. The color map visualizes pointwise geometric
    deviation.}
  \label{fig:ablation_bhc}
\end{figure}

Fig.~\ref{fig:ablation_bhc} shows the visual comparison. When the blind-hole
feature is not assigned a BHC primitive label, the generated polycube fails to
preserve the intended local blind-hole feature, and exhibits substantially
larger geometric deviations near the blind-hole region. In contrast, when the
corresponding cell is assigned the \texttt{BHC+Z} primitive category label, the
generated polycube better preserves the intended local hole-like feature. This
difference is also reflected by the deviation maps. The quantitative comparison
is reported in Tab.~\ref{tab:ablation_dh}. With the introduced BHC primitive,
the average Chamfer distance is reduced from 0.2785 to 0.0292, and the maximum
deviation is reduced from 0.9126 to 0.0842. These results indicate that the BHC
primitive provides an important feature prior for modeling local hole-like
features that do not change the global genus, and substantially improves
polycube generation quality for such cases.

We next analyze the scalability of the proposed automated inference strategy
using representative models shown in
Figs.~\ref{fig:reverse_diffusion}--\ref{fig:partial_results_part2}. Here,
scalability refers to the ability of the method to remain effective when the
primitive set is enlarged, the grid configuration is extended to
$G_{3\times 2\times 2}$, and the corresponding candidate context space grows
substantially. In this setting, direct traversal of all genus-consistent global
contexts becomes increasingly expensive. It is therefore important to examine
how the proposed local context generation and hierarchical verification behave
as structural complexity increases.

Tab.~\ref{tab:scalability} summarizes the average number of assembled global
context candidates, inference time, and verified-candidate rate for models with
different numbers of occupied cells. For each complexity level, the reported
values are averaged over two representative test models. Specifically, the
simple group contains Spline Link and Snow Man, the medium group contains
Squirrel and Face Block, and the complex group contains Egyptian Bust and Relief
Bust. These models are selected from the representative examples shown in
Figs.~\ref{fig:reverse_diffusion}--\ref{fig:partial_results_part2}. For simple
cases with 1--3 occupied cells, the average number of global context candidates
is 27 and the average inference time is about 48~s. For medium cases with 4--6
occupied cells, these values increase to 158 and 280~s, respectively. For more
complex cases with 7 or more occupied cells, the average number of global
context candidates further increases to 510 and the average inference time rises
to about 906~s. Meanwhile, the verified-candidate rate decreases from 25.9\% to
5.1\%. Although the search cost increases nonlinearly with structural
complexity, the results show that the proposed automated pipeline remains
capable of reducing the effective search space and identifying valid solutions
without exhaustive traversal of the full global context space. These results
support our claim that SDDPM improves the scalability of diffusion-based
polycube generation under the expanded primitive set and the enlarged grid
configuration considered in this work.

\begin{table}[pos=htbp]
  \centering
  \caption{Ablation study on the effect of the BHC primitive.}
  \label{tab:ablation_dh}
  \begin{tabular}{lcccc}
    \toprule
    Setting & Local primitive label & Avg. CD ($\downarrow$) & Max. Dev. ($\downarrow$) & Blind-hole feature \\
    \midrule
    Ablation setting & \texttt{Cube}  & 0.2785          & 0.9126          & Not preserved \\
    Proposed setting & \texttt{BHC+Z} & \textbf{0.0292} & \textbf{0.0842} & Preserved \\
   \bottomrule
  \end{tabular}
\end{table}

\begin{table}[pos=htbp]
  \centering
  \caption{Scalability analysis of the automated inference strategy across different levels of structural complexity. The reported values are averaged over two representative models in each complexity group.}
  \label{tab:scalability}
  \begin{tabular}{lccc}
    \toprule
    Problem complexity & Number of global context candidates & Inference time & Verified-candidate rate \\
    (active component count) & (avg.) & (s) & (\%) \\
    \midrule
    Simple (1--3) & 27  & 48  & 25.9 \\
    Medium (4--6) & 158 & 280 & 7.1 \\
    Complex (7+)  & 510 & 906 & 5.1 \\
    \bottomrule
  \end{tabular}
\end{table}

\subsection{All-hex control mesh generation and volumetric spline construction}
Once the polycube structure is generated, it serves as the parametric domain for
all-hex control mesh generation via parametric mapping and adaptive octree
subdivision. We establish a bijective mapping between the input triangular
surface and the polycube boundary surface using the method
in~\cite{tong2026hexopt}, while internal regions are parameterized through
linear interpolation. The all-hex control mesh is then generated through octree subdivision
of the polycube cells, employing an adaptive stopping criterion: if the distance
between a mapped midpoint and its linearly interpolated counterpart falls below
a prescribed threshold, subdivision of that element terminates.

The initial all-hex control mesh may contain distorted elements. We therefore apply the mesh
quality improvement techniques \cite{tong2026hexopt}, including pillowing,
smoothing, and Jacobian-based optimization. Pillowing inserts a boundary layer
to improve mesh quality. Smoothing relocates vertices to improve mesh quality
while preserving geometric features.  Optimization improves the scaled Jacobian
by minimizing an energy that balances geometry fitting and element shape
metrics. By alternately applying these mesh quality improvement techniques, we
obtain an all-hex control mesh with improved element quality as measured by the scaled
Jacobian.

After obtaining an optimized all-hex control mesh, we construct analysis-suitable
volumetric splines using TH-spline3D~\cite{wei17a,yu2020hexgen}. The all-hex control mesh is
treated as the control mesh, and TH-spline3D supports local refinement. The
resulting volumetric splines maintain $C^2$ continuity in regular regions and
$C^0$ continuity around extraordinary points.  Finally, we perform trivariate
B\'{e}zier extraction and export the B\'{e}zier element information via
Hex2Spline~\cite{githexgenhex2spline}, enabling direct integration with IGA
solvers such as ANSYS-DYNA.

\begin{figure}[pos=htbp]
  \centering
  \includegraphics[width=\linewidth]{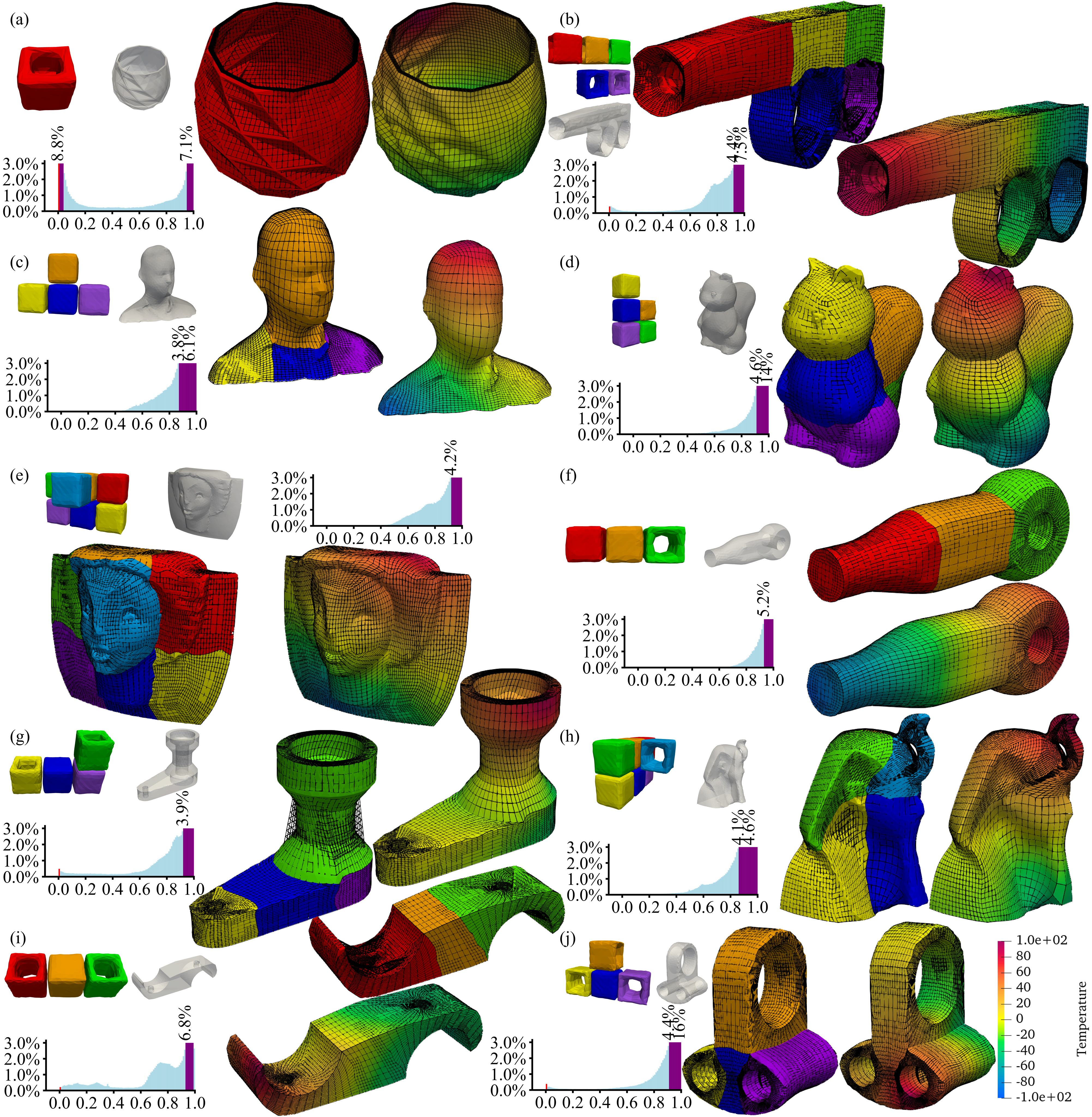}
  \caption{Polycube structures, all-hex control meshes, scaled Jacobian histograms, and volumetric splines with IGA temperature analysis for ten representative models. The red bar in the histograms represents the minimum scaled Jacobian (sometimes the red bar is too short to be seen). }
    \label{fig:context_competition_mesh_quality}
\end{figure}

Fig.~\ref{fig:context_competition_mesh_quality} presents the generated polycube
structures, all-hex control meshes, scaled Jacobian histograms, and volumetric
splines with IGA temperature analysis for ten selected models. These examples
show that the generated polycube structures can be used directly as parametric
domains for all-hex control mesh generation. As shown in Tab.~\ref{tab:statistics}, the
resulting meshes exhibit high quality across a variety of geometries. The scaled
Jacobian histograms show that most hex elements are concentrated in the
high-quality range, while only a small fraction lies near zero. These
distributions indicate that the generated polycube structures provide suitable
parameterization domains for all-hex control mesh generation and largely avoid severe
distortions or inverted elements. Overall, the results confirm that SDDPM can
serve as a reliable front-end for all-hex control mesh generation, followed by volumetric
spline construction and IGA.

\begin{table}[pos=htbp]
\centering
\caption{Statistics of ten tested models in Fig.~\ref{fig:context_competition_mesh_quality}.}
\label{tab:statistics}
\begin{tabular}{ccccc}
\toprule
  Model & \makecell{Input Triangle Mesh\\(vertices, elements)} & \makecell{Deepest\\ Level} & \makecell{Output All-Hex Control Mesh\\(vertices, elements)} & \makecell{Minimum\\ Scaled Jacobian} \\
\midrule
Faceted Pot & (512, 1,020) & 4 & (171,105, 163,840) & 0.02 \\
Triple-Hole Bracket & (2,560, 5,126) & 3 & (89, 454, 81,920) & 0.02 \\
Bust & (2,048, 4,092) & 3 & (140,481, 131,072) & 0.41 \\
Squirrel & (2,560, 5,116) & 3 & (23,137, 20,480) & 0.42 \\
Relief Bust & (3,584, 7,164) & 3 & (32,113, 28,672) & 0.41 \\
Rod & (1,536, 3,072) & 3 & (27,744, 24,576) & 0.66 \\
Cup Bracket & (2,048, 4,096) & 3 & (379,104, 360,448) & 0.02 \\
Egyptian Bust & (3,584, 7,168) & 3 & (346,368, 327,680) & 0.32 \\
Clamp Bracket & (1,536, 3,076) & 3 & (312,543, 294,912) & 0.02 \\
Triple-Hole Vertical Ring Bracket & (2,048, 4,104) & 3 & (450,846, 425,984) & 0.01 \\
\bottomrule
\end{tabular}
\end{table}

\subsection{Limitations}

\label{sec:limitations}
Despite the improvements introduced by SDDPM, several
limitations remain. First, the current method still relies on a fixed
grid configuration, namely $G_{3\times 2\times 2}$, and on a finite
primitive set. Although this design improves representational capacity
compared with the previous $G_{2\times 1}$ setting, it may still be
insufficient for highly complex industrial geometries with richer local
features or more complicated global structures. Extending the current
method to adaptive or hierarchical grids, together with a richer
primitive set, would further improve its flexibility and applicability.

Second, the automated context generation procedure currently depends on a
temporary tetrahedralization and a volumetric partition strategy to construct
local subregions. The robustness of automated inference therefore depends, to
some extent, on the quality of this temporary partition. In addition, the
hierarchical verification module is currently used as a post-generation
validation stage rather than as an explicit constraint during reverse
diffusion. Although GOCC and TCV effectively filter invalid candidates, TCV may
occasionally reject otherwise usable candidates because of localized geometric
irregularities or nonuniform point distributions.

Third, even with the proposed automated context generation strategy,
diffusion-based inference remains more expensive than direct prediction methods,
such as DL-Polycube~\cite{yu_dl-polycube_2025}, because it still requires
iterative reverse diffusion. Finally, this work focuses on polycube-based
all-hex control mesh generation and volumetric spline construction for IGA. The
current pipeline has not yet been extended to other meshing settings, such as
hex-dominant or hybrid mesh generation. Exploring such extensions could further
broaden the applicability of the proposed method.

\section{Conclusions and future work}
\label{sec:conclusion}
In this paper, we propose SDDPM, a diffusion-based method for generating
topology-consistent polycube structures and for constructing IGA-suitable
all-hex control meshes and volumetric splines from CAD surface models. The core
idea is to model the deformation from an input geometry to its polycube
structure as a context-conditioned denoising process, and to improve scalability
and automation by extending our previous DDPM-Polycube pipeline in three
aspects. First, we expand the primitive set from two primitive geometries to
three by introducing a BHC primitive together with its orientation
variants. This extension addresses the challenge that local hole-like features
are not uniquely reflected by the global genus. Second, we move beyond the
restricted $G_{2\times 1}$ setting and adopt the 3D grid configuration
$G_{3\times 2\times 2}$, which provides richer polycube assemblies and improves
representational capacity for more complex geometries. Third, we develop a
genus-guided context generation strategy with a hierarchical verification
module. This strategy unifies automated and user-guided inference under the same
context interface. The hierarchical verification module combines GOCC with TCV,
thereby improving the reliability of both local and global context validation.

Experimental results demonstrate that SDDPM can generate topology- and
label-consistent polycube structures across geometries of different genus. The
generated polycube structures can be used directly as parametric domains for
all-hex control mesh generation, and the resulting meshes support subsequent TH-spline3D
construction and B\'{e}zier extraction for IGA. The results also show that the
proposed pipeline supports both automated and user-guided inference. Although
this paper focuses on automated context generation, user-guided inference is
also supported and can be more efficient when the provided guidance passes the
genus-based consistency check and the hierarchical verification, since it
reduces the number of reverse diffusion trials and candidate verification steps.

There are several directions for future work. The current method still relies on
a fixed grid and a finite primitive set, so extending it to adaptive or
hierarchical grids together with richer primitive categories would further
improve flexibility for highly complex industrial geometries. The automated
context generation procedure currently depends on temporary tetrahedralization
and volumetric partitioning, and its robustness and efficiency could be further
improved. In addition, the hierarchical verification module is currently used
after candidate generation. Incorporating verification information directly into
the reverse diffusion process may reduce search cost and improve inference
efficiency.  Finally, although this work focuses on polycube-based all-hex
control mesh generation and volumetric spline construction, extending
diffusion-based polycube generation to other meshing settings, such as
hex-dominant or hybrid meshes, would further broaden its applicability. By
addressing these challenges, we believe that SDDPM has the potential to further
improve the automation, robustness, and practical applicability of CAD-to-IGA
pipelines.

\section*{Acknowledgements}
Y. Yu, J. Liu, and H. Lou were supported by the National Natural Science
Foundation of China (Grant No. 62302091) and the Shanghai Pujiang Program (Grant
No. 25PJA005).

\bibliographystyle{cas-model2-names}
\bibliography{Hex-Software_edit_parse}
\end{document}